\documentclass[prb, 11pt,letterpaper,superscriptaddress,floatfix,nofootinbib,notitlepage]{revtex4-1}

\usepackage{textcomp}
\usepackage{amsmath}
\usepackage{amsfonts}
\usepackage{amssymb}
\usepackage{graphicx}
\usepackage{siunitx}
\usepackage{caption}
\usepackage{setspace}
\usepackage{xcolor}
\setcitestyle{super}
\usepackage[colorlinks, citecolor={blue}, linkcolor={black}]{hyperref}

\makeatletter
\renewcommand{\@fnsymbol}[1]{%
  \ifcase#1\or
  \TextOrMath\textdagger\dagger\or
  \TextOrMath\textasteriskcentered*\or
  \TextOrMath\textdaggerdbl\ddagger\or
  \TextOrMath\textsection\mathsection\or
  \TextOrMath\textparagraph\mathparagraph\or
  \TextOrMath\textbardbl\|\or
  \TextOrMath{\textasteriskcentered\textasteriskcentered}{**}\or
  \TextOrMath{\textdagger\textdagger}{\dagger\dagger}\or
  \TextOrMath{\textdaggerdbl\textdaggerdbl}{\ddagger\ddagger}\else
  \@ctrerr\fi}
\makeatother

\begin{document}
\raggedbottom
\emergencystretch=2em

\author{Jeong Min Park}
\thanks{These authors contributed equally to this work.}
\affiliation{Joseph Henry Laboratories, Princeton University, Princeton, NJ, USA}
\affiliation{Department of Physics, Princeton University, Princeton, NJ, USA}

\author{Cristian Voinea}
\thanks{These authors contributed equally to this work.}
\affiliation{School of Physics and Astronomy, University of Leeds, Leeds LS2 9JT, UK}
\affiliation{Center for Computational Quantum Physics, Flatiron Institute, New York, NY, USA}

\author{Yen-Chen Tsui}
\thanks{These authors contributed equally to this work.}
\affiliation{Joseph Henry Laboratories, Princeton University, Princeton, NJ, USA}
\affiliation{Department of Physics, Princeton University, Princeton, NJ, USA}

\author{Songyang Pu}
\affiliation{Department of Physics, Washington University in St. Louis, St. Louis, MO, USA}

\author{Kenji Watanabe}
\author{Takashi Taniguchi}
\affiliation{National Institute for Materials Science; Namiki 1-1, Tsukuba, Japan}

\author{Nigel R. Cooper}
\affiliation{Cavendish Laboratory, University of Cambridge, J. J. Thomson Avenue, Cambridge CB3 0US, United Kingdom}

\author{Michael P. Zaletel}
\affiliation{Department of Physics, University of California at Berkeley, Berkeley, CA, USA}

\author{Zlatko Papi{\'c}}
\affiliation{School of Physics and Astronomy, University of Leeds, Leeds LS2 9JT, UK}

\author{Ali Yazdani}
\email{yazdani@princeton.edu}
\affiliation{Joseph Henry Laboratories, Princeton University, Princeton, NJ, USA}
\affiliation{Department of Physics, Princeton University, Princeton, NJ, USA}

\title{Local spectroscopy of anyons bound to charge traps}

\maketitle

\textbf{Fractional quantum Hall states host anyons, emergent quasiparticles with fractional charge and nontrivial exchange statistics\cite{feldman_fractional_2021}. Controlling, trapping, and braiding anyons are central goals for both fundamental physics and topological quantum computation\cite{nayak_non-abelian_2008}. A key step toward such control is understanding how anyons behave when confined in local potentials, where their internal structure can become relevant. Here, we use the scanning tunneling microscopy/spectroscopy (STM/STS) to study the excitation spectrum in integer and fractional quantum Hall states of monolayer graphene near individual charged impurities. In the integer quantum Hall states, the STS spectra show lifting of orbital degeneracy near defects, appearing as a band of discrete energy levels. In fractional states ($\nu=1/3\ \text{and}\ 2/5$), however, we observe an additional energy splitting of the lowest-energy spectral feature that occurs only when the chemical potential lies within a fractional gap and is absent in compressible or integer regimes. We attribute this to many-body configurations of anyons trapped by an impurity potential. Strikingly, numerical calculations show that the splitting requires an anisotropic confining potential, vanishing for a rotationally symmetric trap. The competing multi-anyon states carry nearly identical charge within the core of the potential but differ in how that charge is redistributed at larger radius. Our results establish local tunneling spectroscopy as a direct probe of anyon bound states, providing a key step toward understanding and controlling their behavior in confined geometries relevant for braiding and fusion.}

The fractional quantum Hall (FQH) effect is the canonical realization of topological order in two dimensions, with
ground states whose elementary excitations carry a fraction of the electron charge and obey exchange statistics beyond
those of bosons and fermions\cite{feldman_fractional_2021}. The topological robustness that protects anyonic exchange
phases against local perturbation underlies proposals to use anyons for fault-tolerant quantum
information\cite{feldman_fractional_2021,nayak_non-abelian_2008}. Realizing this prospect requires localizing anyons,
bringing them together, and exchanging them in a controlled way. Each such operation confines quasiparticles to a finite
region, where the smooth, translationally invariant picture of the bulk FQH fluid no longer holds and the internal
many-body structure of a trapped excitation becomes relevant.

Most of what is known about anyons has been inferred from transport measurements. The fractional charge of FQH
quasiparticles was established through shot-noise
measurements\cite{de-picciotto_direct_1998,saminadayar_observation_1997,dolev_observation_2008,venkatachalam_local_2011},
and their fractional statistics have more recently been demonstrated in electronic interferometers and anyon-collision
experiments\cite{halperin_theory_2011,nakamura_aharonovbohm_2019,nakamura_direct_2020,kim_aharonovbohm_2024,werkmeister_anyon_2025,kim_aharonovbohm_2026,bartolomei_fractional_2020}.
These probes treat anyons as essentially point-like objects propagating along edges or through constrictions; they are
not sensitive to how a quasiparticle is assembled from the underlying correlated fluid, nor to how the quasiparticle is
pinned. Even though pinning is precisely what occurs at a charged impurity, an antidot, or an engineered trap, the
regime in which fusion protocols operate, and a local, energy-resolved measurement of an individual trapped
quasiparticle has remained largely out of reach. There is also considerable interest in how anyons interact with one
another, not only to test the process of possible fusion of non-Abelian
anyons\cite{baraban_numerical_2009,prodan_mapping_2009,macaluso_fusion_2019}, but also to explore whether anyons can
create molecular-like states\cite{gattu_molecular_2025-1,xu_dynamics_2025,wang_anyon_2026}.

Scanning tunneling microscopy and spectroscopy (STM/STS) offers the combination of atomic-scale spatial resolution and
spectroscopic access to local excitations needed to address this regime directly. Graphene is a natural platform: its
surface is exposed to the tunnel junction, and its carrier density can be gate-tuned through a sequence of integer and
fractional quantum Hall states. Recent advances in ultra-clean sample fabrication and work function matching between STM
tip and graphene devices has not only made it possible to perform spectroscopy of integer and fractional quantum Hall
gapped phases but also to visualize various broken symmetry
phases\cite{liu_visualizing_2022,hu_high-resolution_2025,tsui_direct_2024}. Critically, in STM measurements, individual
impurities can be identified and addressed, allowing the excitation spectrum to be measured as a function of position
relative to a single trapping potential.

Impurities are expected to result in fundamentally different phenomena in the two quantum Hall states. In integer
quantum Hall (IQH) states, localization is a single-particle effect: the trapping potential simply binds individual
electrons into discrete orbital states. In FQH states, localization is intrinsically many-body, and removing a single
electron from the correlated fluid nucleates a cluster of fractionally charged
quasiholes\cite{bartolomei_fractional_2020,baraban_numerical_2009,prodan_mapping_2009,macaluso_fusion_2019}. Theory has
proposed that tunneling into an FQH fluid near a localized impurity therefore probes many-body excitations whose
structure reflects the fractionalization of the injected charge into anyons\cite{gattu_molecular_2025-1}. An impurity in
the IQH regime yields a single bound state per angular momentum $m$, whereas FQH bound states can appear as a discrete
band of multiple levels at the same $m$\cite{papic_imaging_2018}. This multiplicity is governed by fractional exclusion
statistics, highlighting the intrinsically many-body nature of FQH impurity bound states.

Here we use STM/STS to measure the local excitation spectrum across a sequence of integer and fractional quantum Hall
states of monolayer graphene in the vicinity of individual, sparsely distributed charged impurities. In the integer
regime, the spectra show the expected breaking of orbital degeneracy, consistent with single-particle Landau-level
trapping. In the fractional regime, in contrast, the lowest-energy spectral feature splits: a splitting that appears
only within the fractional gaps and has no counterpart in the integer states. Guided by model calculations, we identify
these features as distinct many-body configurations of anyons bound to the impurity, establishing local tunneling
spectroscopy as a direct probe of the internal many-body structure of anyon bound states.

\textbf{Spectroscopy of the quantum Hall states in graphene}

High-quality monolayer graphene devices were fabricated on hexagonal boron nitride (hBN) substrates with a graphite back
gate, with atomically clean regions exceeding $\SI{200}{\nano\meter}\times\SI{200}{\nano\meter}$ without any detectable
surface adsorbate (Supplementary Information), enabling spectroscopic studies of isolated charge traps. The STM images
of these regions show a high degree of sample homogeneity and the presence of a homogeneous moir\'e superlattice arising
from graphene/hBN misalignment. In the presence of a perpendicular magnetic field, discrete Landau levels (LLs) develop,
and at sufficiently high magnetic field $B=\SI{13.9}{\tesla}$ and millikelvin temperatures
($T_{\mathrm{eff}}=\SI{210}{\milli\kelvin}$), FQH states with large energy gaps can be detected in STM spectroscopy
(Fig. 1a, Supplementary Information).

In monolayer graphene, the zeroth Landau level is fourfold degenerate near charge
neutrality, spanning filling factors from $\nu=-2$ to $\nu=+2$\cite{liu_visualizing_2022, goerbig_electronic_2011, barlas_quantum_2012}, in which the system develops quantum Hall ferromagnetism as well as various FQH states. We focus on
the filling range between $\nu=-2$ to $\nu=-1$, the first quarter of the zeroth Landau level (LL), where a fully spin/valley
polarized flavor quantum Hall ferromagnetic state of a single LL is known to be established at
$\nu=-1$ \cite{liu_visualizing_2022, young_spin_2012, farahi_broken_2023, atteia_skyrmion_2021}. Figure 1a shows the
differential conductance $dI/dV$ as a function of sample bias voltage $V_B$ and gate voltage $V_G$ in this regime. The
incompressible gap corresponds to the regime where spectral features shift monotonically with gate voltage, reflecting
chemical potential tuning, whereas the compressible states correspond to the flat regions where gate voltage primarily
tunes the carrier density\cite{hu_high-resolution_2025}. The spectra exhibit well-defined energy gaps associated with
FQH states that follow the Jain sequence from the $\nu=-2+1/3$ Laughlin state to high-order fractions up to
$\nu=-2+11/23$, confirming the nonperturbative measurement environment with minimal tip-to-sample interactions. Similar
fractional sequences are observed in other filling sectors, including states at $\nu=-1+1/2$ (see Supplementary
Information). Notably, the fractional state at $\nu=-2+1/3$ exhibits an energy gap of approximately
\SI{21.6}{\milli\electronvolt} (gap extracted with previous protocols\cite{hu_high-resolution_2025}) for hole
excitations, which as we show below is comparable to the typical energy scale of impurity-induced defect potentials. The
large magnitude of this gap is essential for resolving local spectroscopic features associated with FQH excitations near
isolated impurities.

\begin{figure}[htbp]
\centering
\includegraphics[width=0.98\textwidth]{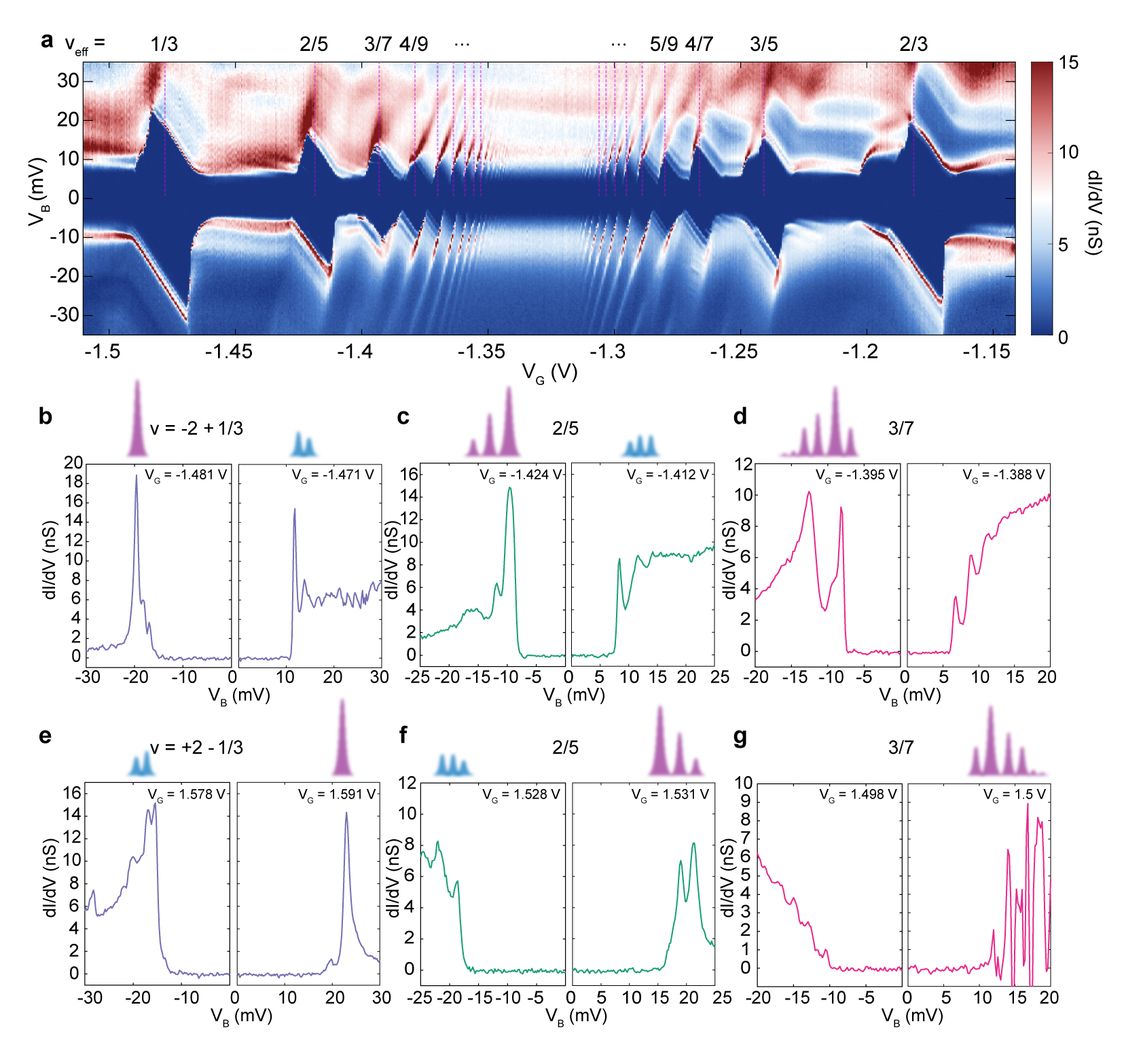}
\caption{\small Scanning tunneling spectroscopy of fractional quantum Hall states in monolayer graphene. (a) Differential conductance $dI/dV$ measured as a function of sample bias $V_B$ and gate voltage $V_G$ over the filling range $-2\leq\nu\leq-1$, acquired at a magnetic field of \SI{13.9}{\tesla}. The spectra reveal well-defined integer and fractional quantum Hall gaps, including a sequence of fractional states consistent with the Jain sequence. (b--d) Tunneling spectra at $\nu=-2+1/3,\,-2+2/5,\,-2+3/7$, respectively, showing the systematic evolution of the bulk LDOS along the Jain sequence. The $\nu=-2+1/3$ spectrum is dominated by a single sharp resonance, while higher-order fractions develop increasingly resolved multi-peak structures consistent with composite-fermion $\Lambda$-level excitations. Schematics show the expected spectral function based on the previous theoretical works\cite{gattu_scanning_2024, pu_fingerprints_2024, yue_electronic_2024}. We note that the calculation for $\nu=3/7$ was not computationally accessible. (e--g) Tunneling spectra at $\nu=+2-1/3,\,+2-2/5,\,+2-3/7$, respectively, showing similar behaviors with the opposite bias voltage.}
\end{figure}

STM tunneling into homogeneous FQH states adds or removes a full electron, which can be understood as probing many-body
composite fermion excitations in the FQH
fluid\cite{jain_reconstructing_2005,gattu_scanning_2024,pu_fingerprints_2024,yue_electronic_2024}. STM spectra (Fig.
1b--d) within the incompressible FQH gaps away from any defects show this evolution across the primary Jain sequence. The
$\nu=-2+1/3$ removal spectrum is dominated by a single sharp resonance, whereas the spectrum for $\nu=-2+2/5$ develops
multiple peaks, consistent with the composite-fermion $\Lambda$-level structure predicted by previous
works\cite{gattu_scanning_2024,pu_fingerprints_2024,yue_electronic_2024}. The number of the resolved peaks qualitatively
agrees well with the calculated spectra, while weaker features may remain unresolved due to the finite resolution. The
multi-peak structure becomes more pronounced in the $\nu=-2+3/7$ spectrum, as expected from the larger number of allowed
composite-fermion configurations at higher-order states in the Jain sequence. These bulk spectra of the FQH states
provide a reference for the impurity measurements below, where a local impurity potential reorganizes the FQH
excitations into bound states of trapped excitations.

\textbf{IQH state near a charge trap}

\begin{figure}[htbp]
\centering
\includegraphics[width=0.85\textwidth]{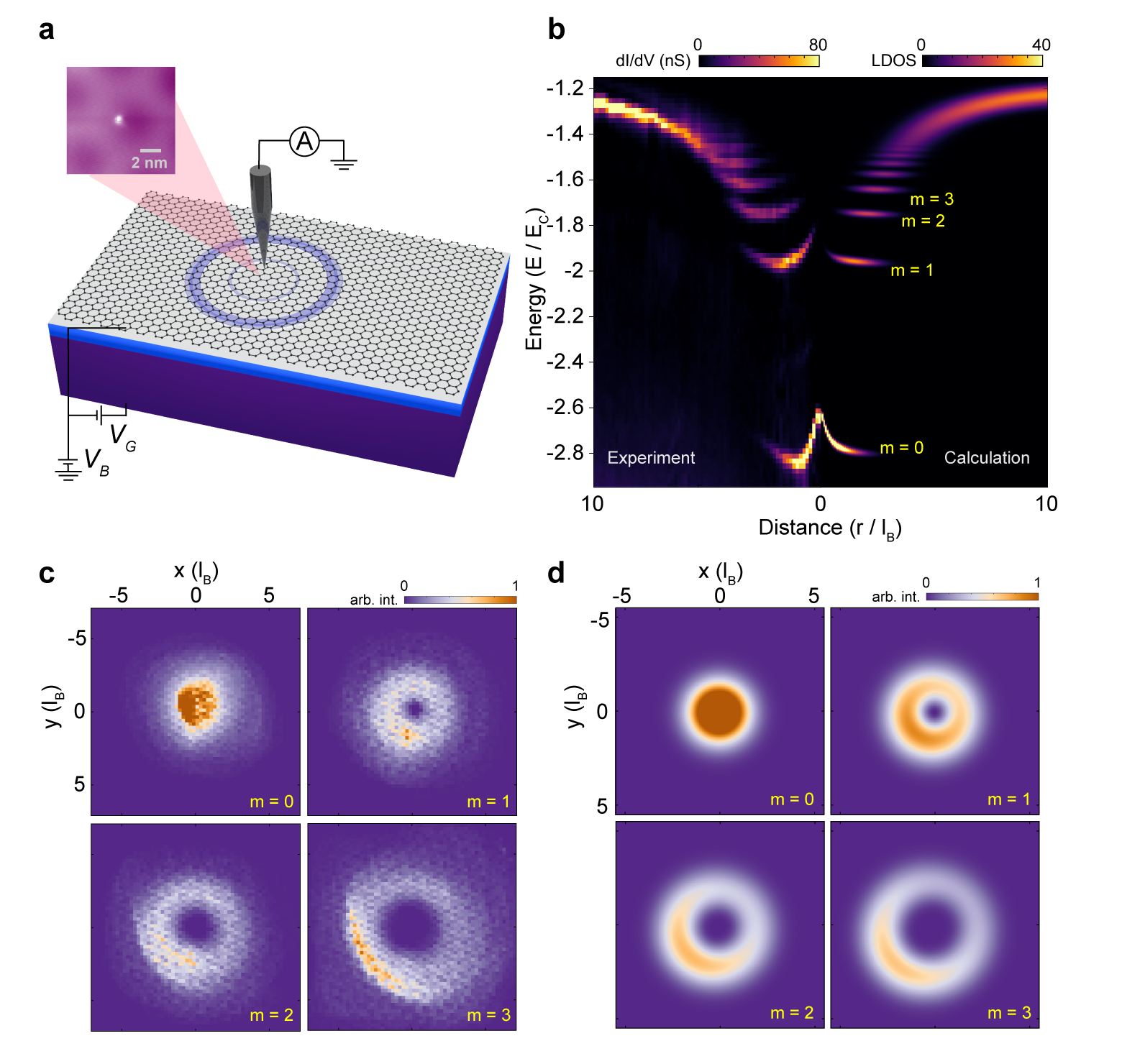}
\caption{\small Angular-momentum dependent degeneracy lifting in integer quantum Hall states near an isolated impurity. (a) Schematic illustration of spatially resolved tunneling spectroscopy across a localized impurity potential. Inset: Topography shows a typical impurity on the surface. (b) Left: Differential conductance $dI/dV$ as a function of sample bias $V_B$ (in units of the Coulomb energy $E_C=e^2/(4\pi\epsilon_0\epsilon_r l_B)$, where $l_B=\sqrt{\hbar/(eB)}$ is the magnetic length, $\epsilon_r\approx4.6$ is the effective dielectric constant) and distance from the impurity for the $\nu=-1$ state, showing the lifting of orbital degeneracy. This impurity, referred to as impurity A, is the main impurity studied in the fractional case in the Main Text; additional data are shown in Supplementary Information. Right: Single-particle calculation of the corresponding spectroscopy. Inclusion of an image charge associated with the STM tip reproduces the bending of the spectral features near the impurity center. (c) Two-dimensional maps of the integrated spectral weight around the impurity for angular-momentum indices $m=0,1,2,3$. The radii of the ring-shaped features agree with the expected values at a magnetic field of \SI{13.9}{\tesla} for the corresponding angular-momentum states. The 2D maps were acquired at another impurity, referred to as impurity B, which exhibits similar linecut behavior (see Supplementary Information). (d) Numerical LDOS calculation can reproduce the measured data well (see Supplementary Information for details).}
\end{figure}

We identify isolated point defects on or beneath an otherwise pristine graphene surface, where they act as localized
electrostatic perturbations. The high sample quality ensures that these defects are spatially well separated, allowing
their effects on the electronic structure to be examined independently. A sharply localized charged defect lifts the
degeneracy of LL orbitals following the screened $1/r$ Coulomb potential (see Fig. 2a inset for a typical impurity). In
the symmetric gauge, each Landau-level orbital has a characteristic radial probability distribution centered at radius
$r_m=\sqrt{2m}\,l_B$, such that the energy shift induced by a defect is determined by the distance between the impurity
and thus the angular momentum wavefunction. Here $m$ is the angular momentum index and $l_B=\sqrt{\hbar/(eB)}$ is the
magnetic length. Orbitals with greater weight near the defect experience larger shifts, whereas orbitals whose
probability density peaks at larger radii remain closer to the unperturbed Landau-level energy. Consequently, a single
LL is split into a discrete set of energy levels corresponding to increasing angular momentum about the center of the
impurity (Fig. 2a).

Our STM measurements directly resolve this angular-momentum-dependent energy splitting of LLs in the integer quantum
Hall regime. Figure 2b shows scanning tunneling spectroscopy acquired across an isolated defect at the $\nu=-1$ gap of
the $N=0$ Landau level, the simplest case with the spin/valley polarized ground states, revealing discrete energy
levels. The observed dispersions are consistent with the theoretically calculated energy shifts, and with previous
measurements\cite{luican-mayer_screening_2014} (see Supplementary Information). The additional bending of the energy
levels near the defect center can be accounted for by electrostatic screening from the STM tip, modeled through an image
charge induced by the tip. Two-dimensional $dI/dV$ maps of the spectroscopic weight further confirm the orbital
character of these states: Fig. 2c shows the integrated spectral density for angular-momentum indices $m=0,1,2,3$,
forming concentric rings whose radii agree with the expected values $r_m=\sqrt{2m}\,l_B$ for a magnetic field of
\SI{13.9}{\tesla}. This behavior is captured by theoretical simulations (Fig. 2d), which include both the effect of the
tip and inter-LL screening (see Supplementary Information). The slightly stronger spectral weight at the lower left
corner of each ring may be due to the nonuniform potential created in the sample, which is also captured by theoretical
calculations.

The close agreement between experiment and theoretical calculation establishes STM spectroscopy near isolated impurities
as a faithful probe of the angular momentum resolved LDOS, providing a calibrated reference for the analysis of FQH
states. We note that we measure both surface and subsurface defects, and do not find qualitative differences (see
Supplementary Information); rather, the behaviors are mostly dependent on the strength of the electrostatic potential
created by the charge traps. For some impurities, angular momentum states $m$ exhibit additional splitting, possibly
related to spin or valley excitations of quantum Hall ferromagnetic states and related to the defect position relative
to the graphene lattice (see Supplementary Information). There are also impurities whose potential is very weak and do
not result in any significant electronic perturbation or impurities that do not appear to trap any charges (see
Supplementary Information), which are not the focus of this work.

\textbf{Charge trap bound resonances in FQH states}

\begin{figure}[htbp]
\centering
\includegraphics[width=0.86\textwidth]{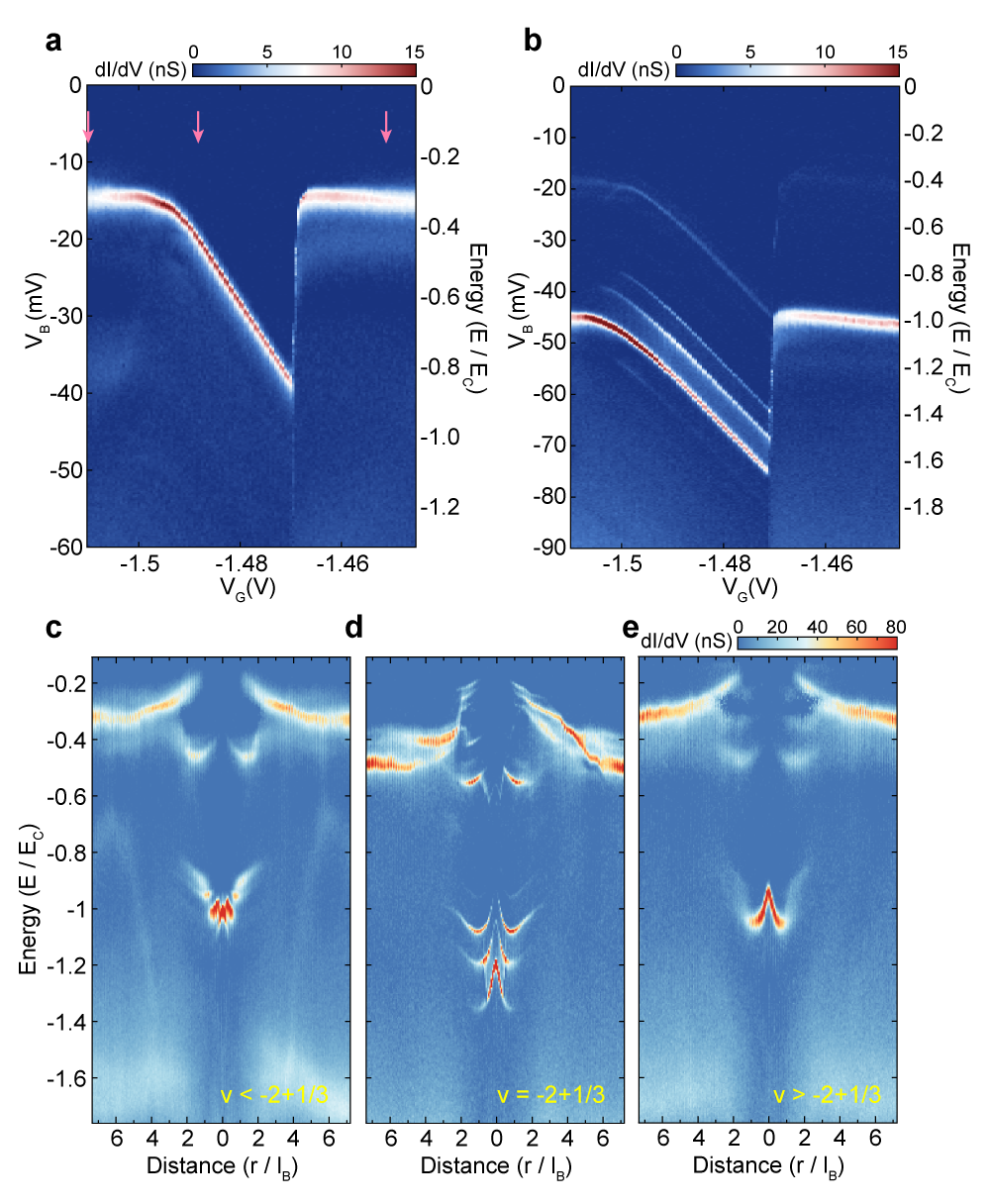}
\caption{\small Energy level splitting at fractional quantum Hall fillings. (a,b) Differential conductance $dI/dV$ as a function of $V_B$ and $V_G$ measured at the $\nu=-2+1/3$ state far from an isolated defect (a) and within its vicinity (b). Arrows indicate the spatial locations where the spectra are acquired. (c--e) Spatially resolved tunneling spectra showing $dI/dV$ as a function of $V_B$ and distance from the defect at gate voltages spanning the $\nu=-2+1/3$ filling. Outside the fractional gap, the lowest-energy spectral feature remains unsplit (c). Upon entering the incompressible fractional regime, the same feature resolves into multiple discrete peaks (d), and reverts to a single peak upon exiting the gap into the compressible regime (e). Corresponding gate voltages are $V_G=-1.51,\,-1.49,\,-1.45\ \mathrm{V}$, respectively, shown with pink arrows in (a). Data were acquired along a line crossing the impurity center for the same impurity as in the IQH state in Fig. 2b (impurity A). Repeating the measurements corresponding to (c--e) with another impurity (impurity C) at a distance of \SI{3}{\nano\meter} from the center shows the same behavior, confirming that the splitting is reproducible and not sensitive to the precise tip position (Supplementary Information). Data for the $\nu=-2+2/5$ state with similar splitting is shown in Supplementary Information.}
\end{figure}

We next examine spectroscopic properties of FQH states in the vicinity of isolated impurities. We focus on the filling
factor $\nu=-2+1/3$, which is the simplest FQH state with partial filling of only one of the fourfold degenerate zeroth
LL starting from $\nu=-2$. Our analysis concentrates on the electron removal spectra (negative-bias; see Fig. 1a and
Supplementary Information for positive-bias addition spectra). Figure 3a shows the tunneling spectrum as a function of
$V_B$ and $V_G$ measured far from any impurities. A well-defined incompressible region is observed, bounded by flat
spectral features that separate the fractional gap from the surrounding compressible states. Within the gap, the local
density of states exhibits a single dominant spectral feature, confirming that there is no splitting due to valley or
spin polarization under this condition. When the same measurement is performed within approximately \SI{2}{\nano\meter}
away from an isolated impurity (Fig. 3b), however, the overall spectral weight shifts to lower energies by about
\SI{30}{\milli\electronvolt} due to the attractive impurity potential, while weak spectral weight remains near the
unperturbed energy. The dominant low-energy feature splits into multiple discrete peaks. Notably, this splitting occurs
only within the incompressible region and is absent in the adjacent compressible regimes.

This behavior is further illustrated in Figs. 3c--e, which show tunneling spectra as a function of energy and distance
across the defect at gate voltages spanning across the $\nu=-2+1/3$ gap. Outside the gap (Fig. 3c), the lowest-energy
feature remains largely continuous, with mostly a single faint spectral feature. Upon entering the incompressible region
(Fig. 3d), a single feature splits into three distinct peaks. As the system is tuned back into the compressible regime
(Fig. 3e), the spectrum reverts to a single peak. An analogous evolution is observed at $\nu=-2+2/5$, where the lowest
energy feature splits into two peaks within the fractional gap, while remaining unsplit outside the gap (Supplementary
Information). Repeating the measurements farther from the defect (in other words, not directly cutting across the center
of the defect), where tip interaction is reduced, yields the same result: the splitting appears exclusively in
fractional incompressible states and only near a charged impurity potential (Supplementary Information).

\textbf{Anyon bound states \& anisotropic potential}

\begin{figure}[htbp]
\centering
\includegraphics[width=0.855\textwidth]{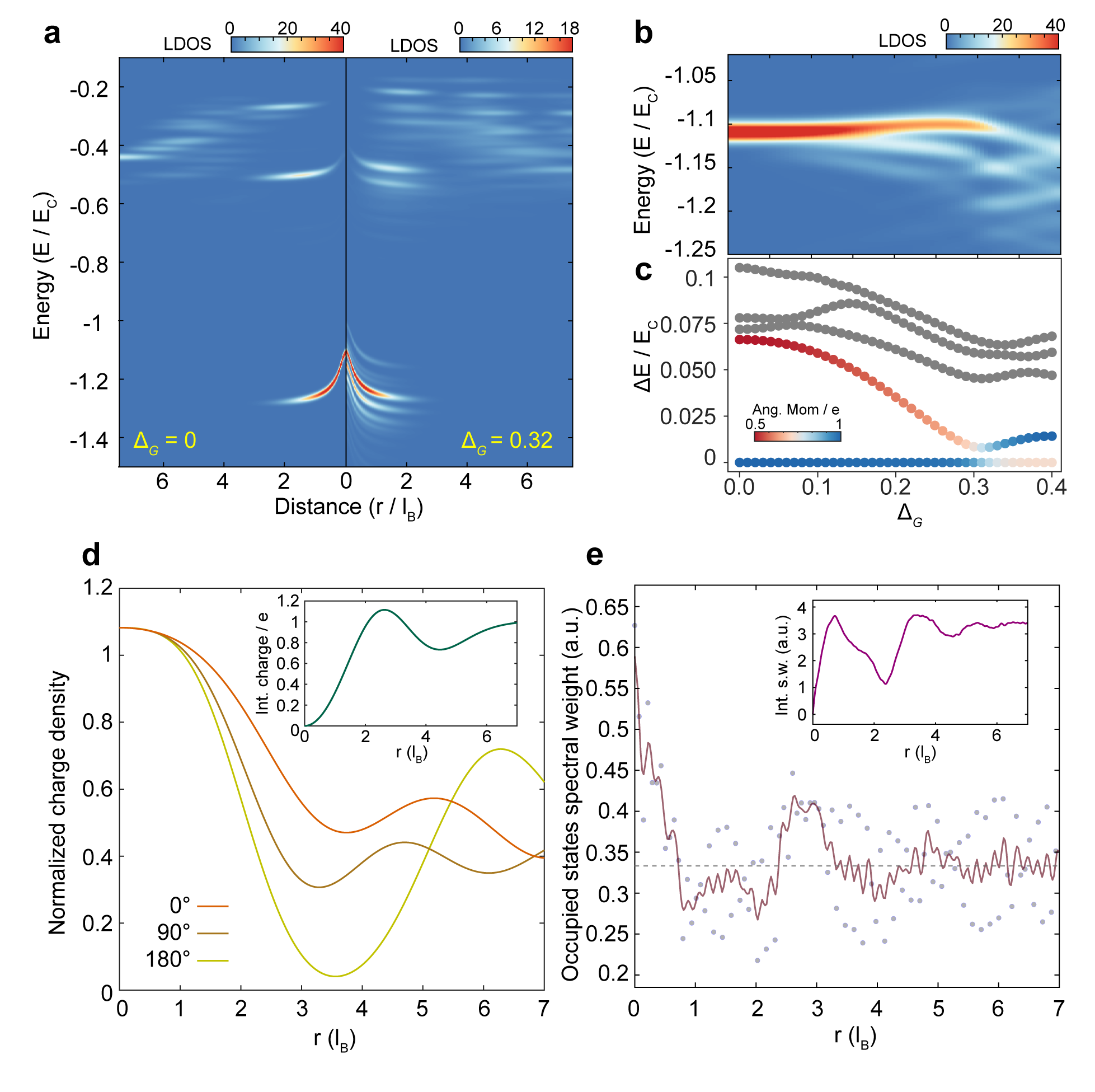}
\caption{\small Numerical simulations for the fractional state with bound anyons and experimental charge density distribution. (a) The STM spectrum is calculated for a system without the potential gradient (left panel) and with the dipolar anisotropic potential of strength $\Delta_G=0.32$ in units of $E_C$ (right panel). Upon turning on the anisotropy, additional resonances appear at the impurity. (b) The evolution of the STM spectrum measured right above the impurity, as the anisotropy $\Delta_G$ is turned on. (c) The evolution of the low-lying energy spectrum before tunneling as a function of the anisotropy strength $\Delta_G$. The color bar shows the average angular momentum of the lowest two eigenstates. The appearance of multiple LDOS peaks is associated with strong hybridization with the first excited state. (d) Theoretically computed charge density profiles, normalized to approach the filling factor $1/3$ far from the anisotropic impurity potential. Different curves correspond to different cuts through the anisotropic potential at angles $0^\circ$, $90^\circ$, and $180^\circ$. All profiles show a localized charge accumulation near the impurity center, followed by oscillations extending over several magnetic lengths. Inset: integrated charge density in a disk of radius $r$ centered at impurity, in units of the electron charge $e$. This shows a trapped charge of approximately $e$, accompanied by a few oscillations further away from the impurity. (e) Integrating the spectral weight of the peaks in the occupied states shown in Fig. 3d, we obtain an experimental charge density distribution for $\nu=-2+1/3$. It is normalized to $1/3$ far from the impurity center. The solid line corresponds to the average over five points. The profile shows a central charge accumulation associated with the trapped quasiparticles, together with charge density oscillations extending away from the impurity center. Inset: excess charge obtained by integrating over $r$ in one direction also shows oscillations in space. We note that the quantitative length scale may depend on several factors, including the detailed impurity potential and tip-induced effects, which have not been examined in detail.}
\end{figure}

The experimental observation of the splitting of the lowest energy impurity-bound resonance in the incompressible gaps
clearly indicates that their explanation requires modeling of low-energy excitation of FQH states near a charged
impurity. We have carried out exact diagonalization calculations in the FQH regime, including perturbing potentials from
both a charged impurity and the STM tip. As described above, the same model captures our IQH results (see Supplementary
Information). For a $\nu=-2+1/3$ Laughlin FQH state in the presence of a strong charged impurity, our calculations show
a single resonance without additional splitting (Fig. 4a). Although this result does not reproduce our experimental
spectra, introducing an anisotropic potential $U_G=\Delta_G\cdot x$, where $\Delta_G$ is the potential strength and $x$ is
the distance along the axis of anisotropy, splits the single resonance near the impurity, matching our experimentally
observed splitting (Fig. 4b). Anisotropy in the local potential could be introduced in many ways. While it naturally
could be created by the tip potential, it can also occur due to the underlying impurity potential. Although the
microscopic origin of the anisotropy remains to be clarified, our results show that it is important for explaining the
multiple splitting of the lowest-energy excitations. The reproducibility of this splitting across different impurities
and tips further indicates that anisotropy is a relevant ingredient for understanding impurity-bound states in the FQH
regime (Supplementary Information).

Studying the energy splitting of the impurity-bound resonance and the pre-tunneling eigenstate evolution as a function
of anisotropy in Figs. 4b--c allows us to understand the microscopic origin of the splitting more clearly. In the
isotropic limit, the LDOS is dominated by a single low-lying impurity resonance; however, once anisotropic potential is
introduced, states with different many-body angular momenta begin to hybridize. Around $\Delta_G\approx0.3$ (in units of
$E_C$, where $E_C$ is the Coulomb energy) in the numerical calculation, the state carrying the dominant impurity
resonance mixes with a nearby excited state, producing an avoided crossing in the low-energy spectrum. The LDOS in Fig.
4b reflects this hybridization: spectral weight that initially appears as a single resonance is redistributed between
the mixed states, resulting in the observed multi-peak structure.

The calculated charge density profile obtained by using the lowest hybridized state in the anisotropic regime, shown in
Fig. 4d, provides a real-space view of the many-body charge rearrangement. The density is strongly enhanced near the
impurity center, consistent with the localization of three $e/3$ anyons, but it does not simply relax monotonically to
the background value. Instead, it exhibits oscillatory charge density modulations extending over several magnetic
lengths away from the center. The integrated excess charge in the inset also shows the oscillatory behavior, indicating
that the anisotropic impurity potential induces an extended redistribution of the charge.

We further compare this theoretical charge rearrangement with the experimentally extracted charge density distribution
for $\nu=-2+1/3$, shown in Fig. 4e. This profile is obtained by integrating the spectral weight of the occupied state
peaks in Fig. 3d and normalizing the density to 1/3 far from the impurity center. The experimental data show the same
qualitative structure as the calculation: a pronounced central charge accumulation associated with trapped
quasiparticles, together with charge density oscillations extending away from the impurity. The excess charge obtained
by integrating the experimental profile in one direction also shows spatial oscillations, as shown in the inset.

Based on the results in Fig. 4, we interpret the low-energy resonance in the absence of an anisotropic potential as the
localization of three $e/3$ anyons. Unlike a single-particle orbital, this impurity-bound state is a correlated
many-body anyon state that can host several nearby internal configurations\cite{wagner_sensing_2026}. By breaking
rotational symmetry, the anisotropic potential mixes many-body configurations with different angular momentum character,
providing a natural mechanism for a single correlated impurity resonance to split into multiple peaks. Intriguingly, the
mixing states with different angular momenta have similar charge distributions in the vicinity of the impurity while
differing on larger length scales (see Supplementary Information). While a full quantitative understanding of the
experimental charge density patterns is left for future studies, our results suggest that local spectroscopy at the
impurity can detect not only the locally trapped charge, but is also sensitive to the longer-distance many-body
properties of the anyonic bound state.

Looking forward, these results motivate several future investigations. Controlled defect engineering using gate-defined
potentials could allow systematic studies of quasiparticle configurations and their response to local anisotropy,
potentially enabling local detection of fusion processes for non-Abelian
anyons\cite{baraban_numerical_2009,prodan_mapping_2009,macaluso_fusion_2019}. Exploring various methods of anyon
confinement beyond that explored here near charge defects, it could also be possible to explore the signatures of anyon
molecules, which have attracted considerable recent theoretical attention and can be tested experimentally using our
approach\cite{gattu_molecular_2025-1,xu_dynamics_2025,wang_anyon_2026}. Extending STM studies to moir\'e materials hosting
fractional Chern insulators\cite{cai_signatures_2023,park_observation_2023,lu_fractional_2024} would allow similar
investigations in zero magnetic field. Finally, combining local STM probes with interferometric device geometries could
provide a powerful hybrid approach for controlling and detecting anyons, bridging real-space imaging and phase-coherent
measurements.

\clearpage

\bibliographystyle{Nature}

\bibliography{references}

\section*{Acknowledgements}

This work was primarily supported by DOE-BES grant and the Gordon and Betty Moore Foundation's EPiQS initiative grants
GBMF9469 to A.Y. Other support for the experimental work was provided by ONR N00012-21-1-2592, NSF-MRSEC through the
Princeton Center for Complex Materials NSF-DMR- 2011750, NSF grant DMR-2312311, NSF grant OMA-2326767, ARO MURI grant
(W911NF-21-2-0147), and ARO grant W911NF261A052. J.M.P. acknowledges Schmidt Science Fellows. N.R.C. was supported by
the EPSRC (Grant No. EP/V062654/1) and by a Simons Investigator Award (Grant No. 511029). C.V. and Z.P. acknowledge
support by the Leverhulme Trust Research Leadership Award RL-2019-015 and EPSRC Grants EP/Z533634/1, UKRI14851. C.V.
acknowledges the Flatiron Institute, a division of the Simons Foundation.

\section*{Author Contributions Statement}

J.M.P. and Y.-C.T. conducted STM measurements and performed data analysis, under the supervision of A.Y. J.M.P.
fabricated the device. C.V. and Z.P. performed theoretical calculations. K.W. and T.T. provided hBN samples. J.M.P and
A.Y. wrote the main manuscript. All authors discussed the results together and contributed to the writing of the
manuscript.

\section*{Correspondence}

Correspondence should be sent to A.Y. (yazdani@princeton.edu).

\section*{Competing Interests Statement}

The authors declare no competing interests.

\section*{Methods}

\subsection{Sample fabrication}

The monolayer graphene device was prepared by assembling an open surface van der Waals heterostructure with a modified
dry pickup technique. A polyvinyl alcohol (PVA) film on transparent tape, mounted on a polydimethylsiloxane (PDMS)
stamp, was used to pick up the exfoliated flakes in the order of monolayer graphene, hBN dielectric layer with a
thickness of $\approx\SI{48}{\nano\meter}$ (corresponding to $\approx7l_B$ at $B=\SI{13.9}{\tesla}$; out-of-plane dielectric constant
$\epsilon_{\perp}\sim3.25$), and a thin graphite flake for bottom gate. The assembled stack was deposited onto a
SiO$_2$/Si chip with pre-patterned Au/Ti electrodes. After release, the PVA layer was removed by repeated cleaning in
HPLC-grade water and solvents, including acetone, isopropyl alcohol, and n-methyl-2-pyrrolidone (NMP), until no visible
residue remained on the graphene surface. The device was then annealed with forming gas at \SI{400}{\celsius}, and
checked with atomic force microscopy (AFM) for cleanliness. The completed device was subsequently introduced into an
ultrahigh-vacuum system and annealed overnight at \SI{475}{\celsius} before STM measurements.

\subsection{STM/STS measurements and data analysis}

Scanning tunneling microscopy and spectroscopy measurements were performed in a home-built dilution refrigerator STM.
The mixing chamber temperature was approximately \SI{20}{\milli\kelvin}, with the effective electron temperature around
\SI{210}{\milli\kelvin} (calibrated using an Al (100) reference surface). All measurements were performed at
perpendicular magnetic field of \SI{13.9}{\tesla}. Tungsten tips were conditioned on a Cu (111) surface before
measurements. During the experiments, the STM tip was grounded, while sample bias $V_B$ was applied to MLG. To maintain
the electrostatic potential difference between MLG and back gate and control the gate voltage by $V_G$, combined bias
and gate voltages $V_B+V_G$ were applied to the back gate.

Differential conductance spectra were acquired using standard lock-in detection of differential conductance $dI/dV$.
Tunneling spectroscopy was typically performed with a tunneling set point in the range of $V_B=100\text{--}400\,\mathrm{mV}\ \text{and}\ I=1\text{--}3\,\mathrm{nA}$, with \SI{1}{\milli\volt} a.c. modulation at \SI{712.9}{\hertz}. Spatially resolved spectroscopy was
performed in constant height mode using either one-dimensional linecuts or two-dimensional spectroscopic grids. Before
each measurement, the local sample plane was carefully levelled so that the tip followed a flat trajectory with minimal
variation in tip-sample distance. The feedback loop was first stabilized at a chosen reference condition in $V_B$ and
$V_G$, then turned off. With the feedback disabled, $V_B$ was swept in the desired spectroscopic range for the linecut
or grid measurement. Measurements were taken only after sufficient settling time to minimize lateral drift. To construct
the angular-momentum-resolved 2D maps in the IQH regime, we first identified the $m=0,1,2,3$ resonances from
energy-resolved linecuts. Because the resonances are broadened and shifted by tip-induced bending, the spatial area
associated with each $m$ state was selected separately at each energy by comparison with the corresponding linecut
feature. The spectral weight within these energy-dependent spatial regions was then integrated over the relevant energy
range. For charge distribution in Fig. 4, we integrated the spectral weight of the occupied state peaks by selecting the
main peak ranges and integrating the nonnegative signals.

\end{document}

\title{
Supplementary Information for ``Local spectroscopy of anyons bound to charge traps''
}

%\author{...}
% \affiliation{School of Physics and Astronomy, University of Leeds, Leeds LS2 9JT, UK}

\begin{abstract}
In this Supplementary Information, we provide additional data supporting the results in the main text, with the detailed description of the theoretical model and exact diagonalization results for integer and fractional quantum Hall states.
\end{abstract}

\maketitle

\section{Details of numerical simulations}

In this section, we provide details of our numerical calculations based on exact diagonalization, before presenting the results for the integer quantum Hall (IQH) state at filling factor $\nu=1$, which serves as a benchmark for our model, and the new features related to anisotropy-induced splitting of STM signal in the fractional quantum Hall (FQH) state at filling $\nu=1/3$. 

\subsection{The model}

Our numerical calculations are performed in the spherical geometry~\cite{Haldane83}. We place the
two-dimensional electron gas on the surface of a sphere with a magnetic monopole
of strength \(2Q\) at the center.  The sphere radius is
\(R=\sqrt{Q}\ell_B\), where \(\ell_B\) is the magnetic length.  Quantum Hall
states are obtained by relating the number of flux quanta to the number of
electrons $N_e$ through \(2Q = \nu^{-1}N_e-\mathcal{S}\), where $\nu$ is the filling factor in the thermodynamic limit and \(\mathcal{S}\) is the
so-called Wen-Zee shift~\cite{Wen92}.  For the IQH  case
we use \(2Q=N_e-1\), giving a filled lowest Landau (LLL) level in a single spin/valley
branch.  For the FQH state at \(\nu=1/3\), we use \(2Q=3(N_e\pm1)-3\),
corresponding to the Laughlin state with an additional electron or hole.  This
choice is useful for strong impurities that can trap a full electron or hole charge near
the impurity, because it avoids forcing a compensating charge depletion far
away from the impurity. Note that only the total extra charge is fixed but not its spatial distribution; the latter is determined self-consistently by the system to minimize the total energy, which involves both the interactions and the impurity potential.

The STM signal is characterized by the local density of states (LDOS) for adding
or removing an electron~\cite{He93b,Haussmann96,Jain05,Papic18,Xiaomeng2022,Farahi2023,Pu2023,Gattu2023,Yue2024},
\begin{align}
\mathrm{LDOS}(\omega,\theta,\varphi)
&=
\sum_a
\delta\!\left(\omega-(E_a-E_\Omega)\right)
\left|
\langle a|\hat{\Psi}^{(\dagger)}(\theta,\varphi)|\Omega\rangle
\right|^2 .
\label{eq:LDOSdef}
\end{align}
Here \(|\Omega\rangle\) is the \(N_e\)-electron quantum Hall ground state before the tunneling process,
 \(\hat{\Psi}(\theta,\varphi)=
\sum_{m=-Q}^{Q}Y_{QQm}(\theta,\varphi)\hat c_m\) is the LLL-projected electron
field operator on the sphere, expressed in terms of monopole harmonics $Y_{QQm}$,  \(|a\rangle\) are eigenstates in the particle-number sector after
tunneling ($N_e\pm 1$).  The dagger is used for electron addition; without it, the expression
gives the electron-removal LDOS.

Because the electrostatic potential of the tip depends on its position, the
post-tunneling states \(|a\rangle\) generally depend on spherical angles \((\theta,\varphi)\).
For subsequent visualizations, we broaden the delta function with a Gaussian of width
\(\sigma\):
\begin{align}
\mathrm{LDOS}(\omega,\theta,\varphi)
&=
\frac{1}{\sigma\sqrt{2\pi}}
\sum_a
e^{-(\omega-(E_a-E_\Omega))^2/(2\sigma^2)}
\left|
\sum_{m=-Q}^{Q}
Y_{QQm}(\theta,\varphi)
\langle a|\hat c_m|\Omega\rangle
\right|^2 .
\end{align}

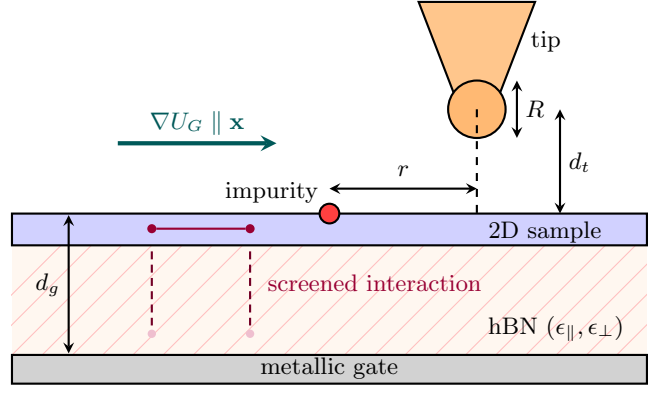
\begin{figure}[!tb]
\centering
\begin{tikzpicture}[x=1cm,y=1cm,>=stealth]

  % hBN dielectric region between sample and gate
  \begin{scope}
    \clip (-4.2,-1.45) rectangle (4.2,0.0);
    \fill[orange!6] (-4.2,-1.45) rectangle (4.2,0.0);
    \foreach \x in {-5.0,-4.6,...,4.6} {
      \draw[red!25,thin] (\x,-1.45) -- ++(1.45,1.45);
    }
  \end{scope}
  \node[black] at (3,-1.1)
    {\small hBN $(\epsilon_{\parallel},\epsilon_{\perp})$};
    
  \fill[blue!18] (-4.2,0.0) rectangle (4.2,0.42);
  \fill[gray!35] (-4.2,-1.45) rectangle (4.2,-1.83);
  \draw[thick] (-4.2,0.0) rectangle (4.2,0.42);
  \draw[thick] (-4.2,-1.45) rectangle (4.2,-1.83);
  \node at (2.85,0.18) {\small 2D sample};
  \node at (0,-1.62) {\small metallic gate};

  \fill[red!75] (0,0.42) circle (0.13);
  \draw[thick] (0,0.42) circle (0.13);
  \node[above left=1pt] at (0,0.42) {\small impurity};

  \fill[orange!45] (1.18,3.22) -- (2.72,3.22) -- (2.25,2.00) -- (1.65,2.00) -- cycle;
  \draw[thick] (1.65,2.00) -- (1.18,3.22) -- (2.72,3.22) -- (2.25,2.00);
  \fill[orange!55] (1.95,1.80) circle (0.38);
  \draw[thick] (1.95,1.80) circle (0.38);
  \node[right] at (2.55,2.70) {\small tip};
  \draw[<->,thick] (2.48,1.42) -- (2.48,2.18);
  \node[right] at (2.48,1.80) {\small $R$};

  \draw[dashed,thick] (1.95,1.80) -- (1.95,0.42);
  \draw[<->,thick] (3.05,0.42) -- (3.05,1.80);
  \node[right] at (3.05,1.10) {\small $d_t$};

  \draw[<->,thick] (-3.45,0.40) -- (-3.45,-1.43);
  \node[left] at (-3.45,-0.52) {\small $d_g$};

  \draw[<->,thick] (0,0.75) -- (1.95,0.75);
  \node[above] at (0.98,0.80) {\small $r$};

  \draw[->,ultra thick,teal!70!black] (-2.8,1.35) -- (-0.7,1.35);
  \node[above,teal!70!black] at (-1.75,1.35) {\small $\nabla U_G \parallel \mathbf{x}$};

  \fill[purple!80!black] (-2.35,0.22) circle (0.06);
  \fill[purple!80!black] (-1.05,0.22) circle (0.06);
  \draw[thick,purple!80!black] (-2.29,0.22) -- (-1.11,0.22);
  \draw[densely dashed,purple!60!black,thick] (-2.35,-0.08) -- (-2.35,-1.06);
  \draw[densely dashed,purple!60!black,thick] (-1.05,-0.08) -- (-1.05,-1.06);
  \fill[purple!25] (-2.35,-1.17) circle (0.055);
  \fill[purple!25] (-1.05,-1.17) circle (0.055);
  \node[purple!80!black] at (0.6,-0.5) {\small screened interaction};

\end{tikzpicture}
\caption{Model geometry.  The hBN layer with dielectric constants \((\epsilon_\parallel, \epsilon_\perp)\) and the metallic gate at a distance \(d_g\) below the
2D sample screen Coulomb interactions.  A localized charged impurity sits
near the sample surface.  A metallic tip of radius \(R\) probes the LDOS at
height \(d_t\) and lateral displacement \(r\) from the impurity.  We
also include an in-plane gradient potential \(U_G\) to model broken rotational
symmetry in the experimental STM spectra.}
\label{fig:setup_model}
\end{figure}

The model geometry is summarized in Figure~\ref{fig:setup_model}. 
The metallic gate at distance \(d_g\), together with the hBN dielectric
environment, gives the screened electron-electron interaction
\begin{equation}
V_C(q)
=
\frac{4\pi}{q}
\frac{1}{1+\epsilon\coth(\beta q d_g)} ,
\label{eq:coulomb_screened}
\end{equation}
where \(q\) is the magntude of the in-plane momentum.  The dielectric parameters are
\(\epsilon=\sqrt{\epsilon_\parallel\epsilon_\perp}\) and
\(\beta=\sqrt{\epsilon_\parallel/\epsilon_\perp}\), with
\(\epsilon_\parallel\simeq6.6\) and \(\epsilon_\perp\simeq3.25\).  Unless
specified otherwise, the reference gate distance is \(d_g=7\,\ell_B\).
Inter-LL screening is included through the Random Phase Approximation
(RPA)~\cite{Shizuya2007},
\begin{equation}
V_{\mathrm{RPA}}(q)=\frac{V_0(q)}{1-V_0(q)\Pi(q)},
\label{eq:rpa_convention}
\end{equation}
where the graphene fine-structure constant is taken to be \(\alpha_G=1.85\).

An impurity of charge $Q_I$ at distance \(d_i\) from the sample induces the one-body potential
\begin{equation}
U_I(q)
=
\frac{4\pi Q_I}{q}
\frac{\sinh[\beta q(d_g-d_i)]\,\sinh^{-1}(\beta qd_g)}
{1+\epsilon\coth(\beta qd_g)} .
\label{eq:impurity_screened}
\end{equation}
In the experimental regime of interest,
the impurity is very close to the 2D sample, hence we will work in the limit \(d_i=0\,\ell_B\) unless otherwise specified.

The metallic STM tip is modeled as a sphere of radius \(R=0.5\,\mathrm{nm}\). The complete solution for the tip, impurity, gate and hBN ensemble does not have a closed form, so we will use the first-order approximation where the radius $R$ of the sphere is taken to be smaller than all other length scales. In this limit, the potential created by the impurity across the entire tip surface is constant, and the resulting induced charge on the tip is
\begin{equation}
    Q_T(r) = -R\,U_I(r, z=d_t) = -R \int \frac{\mathrm{d}^2 q}{(2\pi)^2}  U_I(q,z=0) e^{-q d_t}e^{-i \mathbf{q} \cdot \mathbf{r}} \, ,
\end{equation}
where the Fourier transform of the impurity potential at the sample location $U_I(q,z{=}0)$ takes the form of Eq.~(\ref{eq:impurity_screened}) and is valid for any impurity distance $d_i$.
This charge is found numerically for each position of the tip and approximately placed at the center of the sphere; the resulting tip potential at the sample surface is given by
\begin{equation}
U_T(r,q)
=
\frac{4\pi Q_T(r)}{q}
\frac{\exp(-q d_t)}{1+\epsilon\coth(\beta qd_g)} \,.
\label{eq:tip_screened}
\end{equation}
This image-charge description captures the main qualitative effect of the tip:
as the tip approaches the impurity, it bends the local LDOS resonance pattern.
Unless specified otherwise, we use the experimentally relevant value
\(d_t=0.2\,\ell_B\).  With the magnetic field
\(B=13.9\,\mathrm T\), the magnetic length 
\(\ell_B=25.66\,\mathrm{nm}/\sqrt{B[\mathrm T]}
\approx 6.87\,\mathrm{nm}\) and the physical tip radius corresponds to
\(R/\ell_B \approx 0.073\).  

Finally, the experimental results in the main text show that the spatial density maps of the prominent LDOS peaks are strongly anisotropic. The anisotropy profile is consistent with a dipolar deformation, which we model phenomenologically by including a dipolar potential gradient; on the sphere, this takes the form $U_G (\theta, \varphi) = \Delta_G \sin\theta \cos \varphi$, with a near-impurity expansion in planar geometry of $U_G \sim x$. Its dipolar nature becomes apparent in the angular momentum decomposition
\begin{equation}\label{eq:gradient}
U_G = \Delta_G \, \frac{1}{2(Q+1)}( L_+ + L_-),   
\end{equation}
where $L_{\pm}$ are the angular momentum ladder operators. 
%The matrix elements are
%\begin{equation}\label{eq:imp_matel}
%\langle Q,m'|U_G|Q,m\rangle
%=
%\frac{1}{2(Q+1)}\sqrt{(Q-m)(Q+m+1)}\,\delta_{m',m+1}
%+
%\frac{1}{2(Q+1)}\sqrt{(Q+m)(Q-m+1)}\,\delta_{m',m-1}.    
%\end{equation}
$U_G$ is not diagonal in the LLL basis, hence it breaks the conservation of $L_z = \sum_m m c_m^\dagger c_m$. Nevertheless, one can show that the matrix elements $\langle Q,m'|U_G|Q,m\rangle \propto \delta_{m',m\pm 1}$. Hence, $U_G$ can only couple states whose $L_z$ values differ by $1$ and therefore gives rise to the selection rule $\Delta L_z = \pm 1$. 

To avoid edge effects, we solve the many-body problem on the sphere, while the screened interactions, impurity potential, and tip potential are evaluated in planar geometry; the spherical matrix elements converge to these planar expressions in the large-$R$ limit.
All one- and two-body potentials are projected onto the LLL.  In practice, the
impurity and tip potentials are tabulated in momentum space and converted into
LLL matrix elements, while the electron-electron interaction is encoded through
screened Haldane pseudopotentials.  The impurity is placed at one pole of the
sphere.  The tip is moved during the LDOS calculation by rotating the one-body
potential with Wigner \(D\)-matrices at each tip position.  Below, we quote all distances in units of \(\ell_B\) and energies in units of the Coulomb scale
\(E_C=e^2/(\epsilon\ell_B)=e^2/(\sqrt{\epsilon_\parallel\epsilon_\perp}\ell_B)\). 
The inclusion  of the tip and gradient potential considerably increases the complexity by breaking the $L_z$-symmetry and changing the Hamiltonian at each position of the tip. Furthermore, since the main features above the impurity appear at large negative bias, the entire eigenspectrum needs to be obtained in order to satisfy the LDOS sum rule, rather than a handful of low-lying states. Consequently, the numerical simulations are restricted to smaller particle numbers compared to previous studies of low-lying spectra with conserved $L_z$.

The LDOS energy axis includes a correction for the neutralizing electrostatic background, following the convention of
Refs.~\cite{Morf02,Balram20b}.  From the interaction pseudopotentials \(V_m\), we compute the average charging energy
\begin{equation}\label{eq:eps_ch}
   \epsilon_{\mathrm{ch}}=\sum_{m=0}^{2Q}V_m\frac{(4Q-2m+1)}{(2Q+1)^2},
\end{equation}
and the total energy $\mathcal E$ is obtained by subtracting a term \( \tilde N^2\epsilon_{\mathrm{ch}}/2\), where $\tilde N^2 \equiv N_e^2$ for the vacuum and $\tilde N^2 \equiv (N_e\pm 1)^2-1$ for the charged sectors.  Therefore, the total LDOS energies are
$\omega_a^\pm = \pm ({\mathcal E}_{N_e\pm 1,a}-{\mathcal E}_{N_e,0})$. 
Finally, it is customary to measure the energies relative to the chemical potential, 
\(\mu=(\mathcal E_{N_e+1,0}-\mathcal E_{N_e-1,0})/2\) and plot
\(E^\pm=\omega^\pm-\mu\), and we will follow this practice for the FQH case. On the other hand, the IQH case is special for the numerics because we do not have access to $\omega^+$ since it is not possible to add an electron to the fully-filled LLL. In this case, to match theory with experiment, we treat the overall reference energy scale as a fitting parameter, as explained below. 

\subsection{Integer quantum Hall case}

\begin{figure}[!tbp]
\centering
\includegraphics[width=0.31\textwidth]{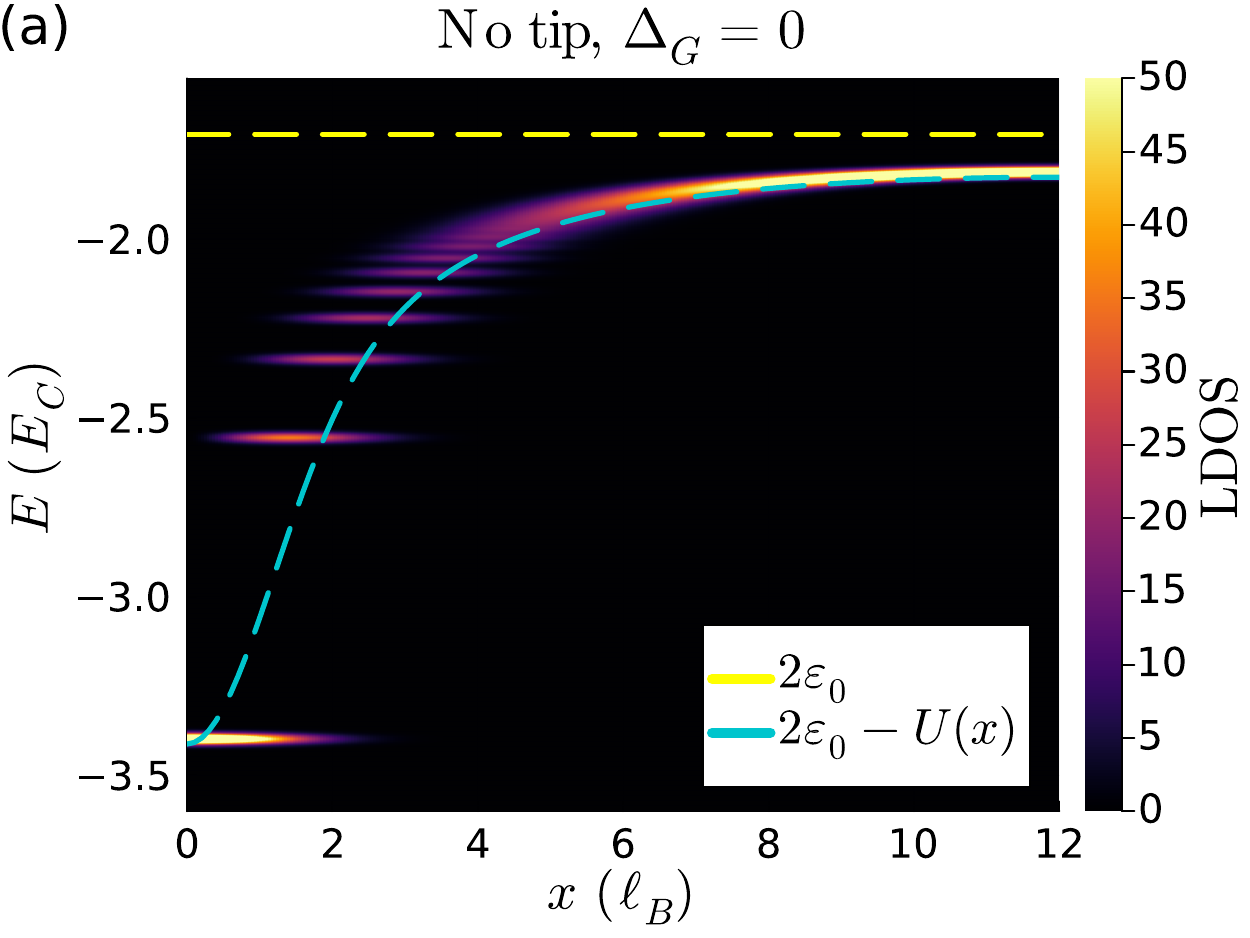}
\includegraphics[width=0.31\textwidth]{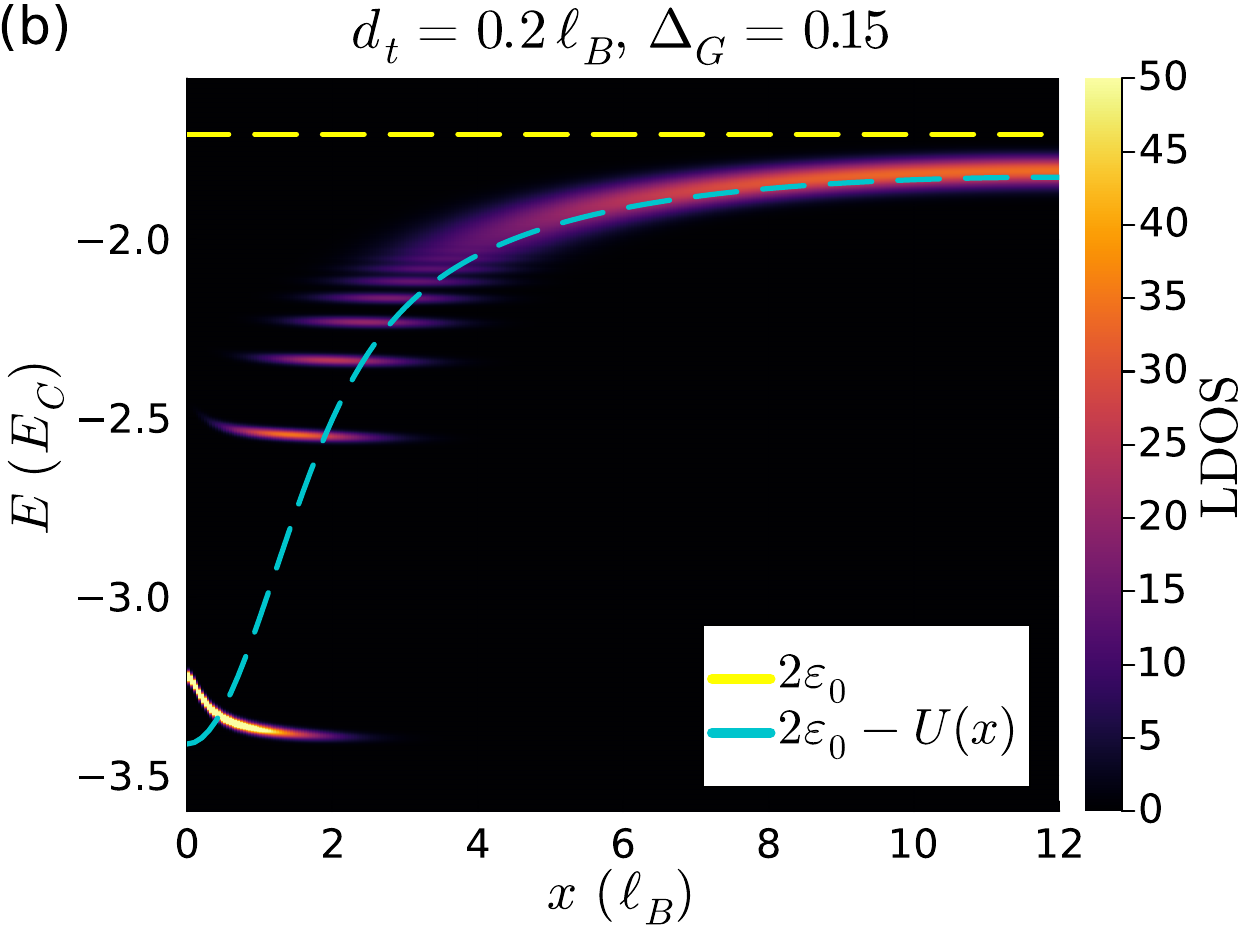}
\includegraphics[width=0.31\textwidth]{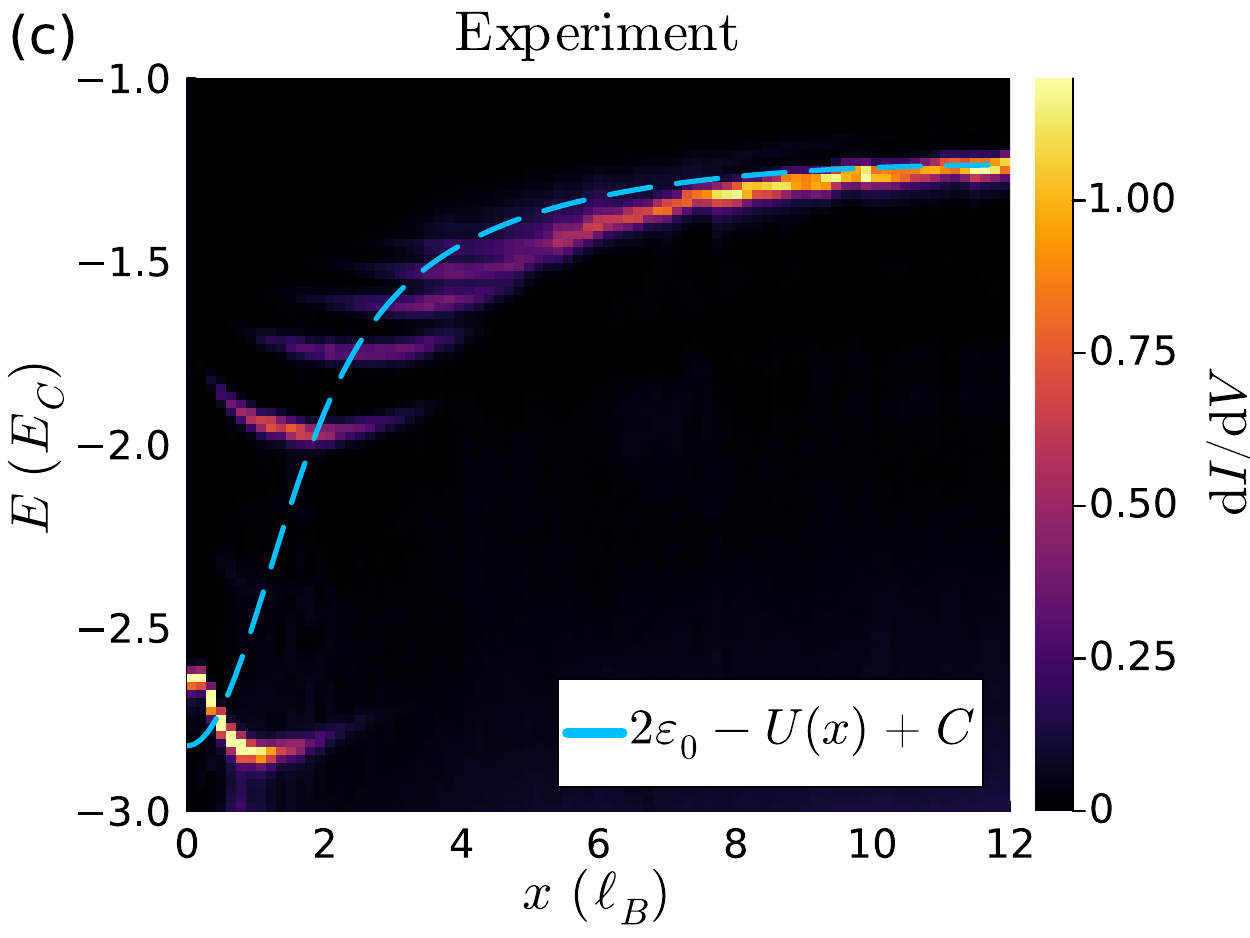}
\caption{
(a),(b): LDOS for IQH state with \(N_e=30\), \(2Q=29\), \(d_i=0\,\ell_B\), \(Q_I=1\),
\(d_g=7\,\ell_B\), and RPA-screened Coulomb interaction.  Panel (a) excludes the scanning tip
and linear gradient field; (b) includes both with \(d_t=0.2\,\ell_B\) and
\(\Delta_G=0.15\).  The finite
\(d_t\) term bends the LDOS peaks most visibly near the impurity
and only weakly at larger \(x\).  By contrast, \(\Delta_G\) is the symmetry
breaking field responsible for the anisotropic LDOS peak density profiles
discussed in the main text.  Axes are lateral position \(x/\ell_B=\sqrt Q\,\theta\) and removal
energy in units of \(E_C\); color is LDOS intensity. Dashed lines are analytical predictions, see text for details. Panel (c) shows the experimental data fitted using the analytical predictions and the single fitted constant offset \(C = 0.59\). The LDOS peaks in panels (a)-(b) are artificially broadened in the energy direction to increase visibility.
}
\label{fig:iqh-local-impurity}
\end{figure}

\Cref{fig:iqh-local-impurity}(a) shows the LDOS removal spectrum of a filled LLL in the presence of a charged impurity.  The vertical axis is the removal energy after subtracting the neutralizing
background, Eq.~(\ref{eq:eps_ch}). In the absence of any local potential, the occupied branch would asymptote towards
\(2\epsilon_0^{\mathrm{clean}} \approx -1.706\,E_C\) (yellow dashed line), which represents (twice) the exchange energy per particle of a filled LLL for our interaction model. (The screening induces a significant correction to the IQH exchange energy, which takes the familiar value \(\epsilon_0^{\mathrm{bare}} = - \sqrt{\pi/8}E_C \approx -1.25\,E_C\) for the bare Coulomb potential.)  Without the impurity, the LDOS spectrum would be perfectly flat and pinned to the energy \(2\epsilon_0^{\mathrm{clean}}\).

The charged impurity shifts the LDOS locally.  The relevant local energy estimate is given by
\begin{equation}\label{eq:impbaseline}
  E_{\mathrm{loc}}^{(-)}(x)\simeq2\epsilon_0^{\mathrm{clean}}
-U_{\mathrm{loc}}(x), \quad U_{\mathrm{loc}}(x) = \sum_j w_j(\theta)U_{I,j}, \quad   w_j(\theta)=\binom{2Q}{j}[\cos^2(\theta/2)]^{2Q-j}
[\sin^2(\theta/2)]^j,
\end{equation}
where $U_{I,j}$ is the impurity matrix element for the $j$th LLL orbital. Thus the no-tip LDOS branch is pulled down in energy near the impurity
and returns toward the pristine exchange value as the local impurity potential
decays, following the blue dashed line in \Cref{fig:iqh-local-impurity}(a). 

\Cref{fig:iqh-local-impurity}(b) then adds the scanning tip potential in Eq.~(\ref{eq:tip_screened}).  Note that the
tip is not a passive readout: at finite \(d_t\) it contributes a
position-dependent one-body potential, so moving the tip changes the Hamiltonian
whose LDOS is being measured.  This tip-induced perturbation bends the LDOS peak
energies most strongly when the tip is close to the charged impurity, where the
image charge is largest, and becomes weak farther away.  

On the other hand, the experimental results in the main text also show that the density maps of the prominent LDOS peaks are markedly anisotropic in space, which we can capture using the gradient potential $U_G$ in Eq.~(\ref{eq:gradient}). 
The anisotropy in experimental density maps is consistent with a dominant dipolar component. In the main text, the value $\Delta_G \approx 0.15\, E_C$ was shown to reproduce the experimentally observed breaking of rotational symmetry. We emphasize that this is an additional effect on top of the tip-induced bending that we see in 
\Cref{fig:iqh-local-impurity}(b).

Putting it all together, \Cref{fig:iqh-local-impurity}(c) demonstrates a quantitative match between theory and experimental LDOS.  We keep the theory parameters fixed to the
\(Q_I=1\), \(d_g=7\,\ell_B\), \(d_t=0.2\,\ell_B\), and
\(\Delta_G=0.15\), and fit LDOS to the form 
\begin{equation}\label{eq:iqh_fit}
  E(x)=2\epsilon_0^{\mathrm{clean}} - U_{\mathrm{loc}}(x)+C,  
\end{equation}
which is the same as Eq.~(\ref{eq:impbaseline}) up to the constant offset
\(C\), which fixes the ``zero voltage'' in the numerical data.  The fit  gives \(C=0.59\,E_C\). After fixing the reference energy $C$, \Cref{fig:iqh-local-impurity}(c) shows a generally good agreement between theory and experiment for realistic values of other parameters, including the energies of the lowest LDOS peaks and their spatial profiles close to the impurity, as demonstrated in the main text.

\subsection{Fractional $\nu=1/3$ case}

While increasing the impurity strength cannot rearrange the charge at integer filling due to the Pauli exclusion principle, this is possible at fractional filling. Consequently, the structure of LDOS is strongly affected by both the strength of the impurity and the breaking of rotational symmetry. Below we first demonstrate the effect of these in our numerical simulations of LDOS at $\nu=1/3$, and then discuss the mechanism behind the observed LDOS splitting.  

\subsubsection{Weak and strong impurity regimes}

\begin{figure}[!b]
\centering
\includegraphics[width=0.32\textwidth]{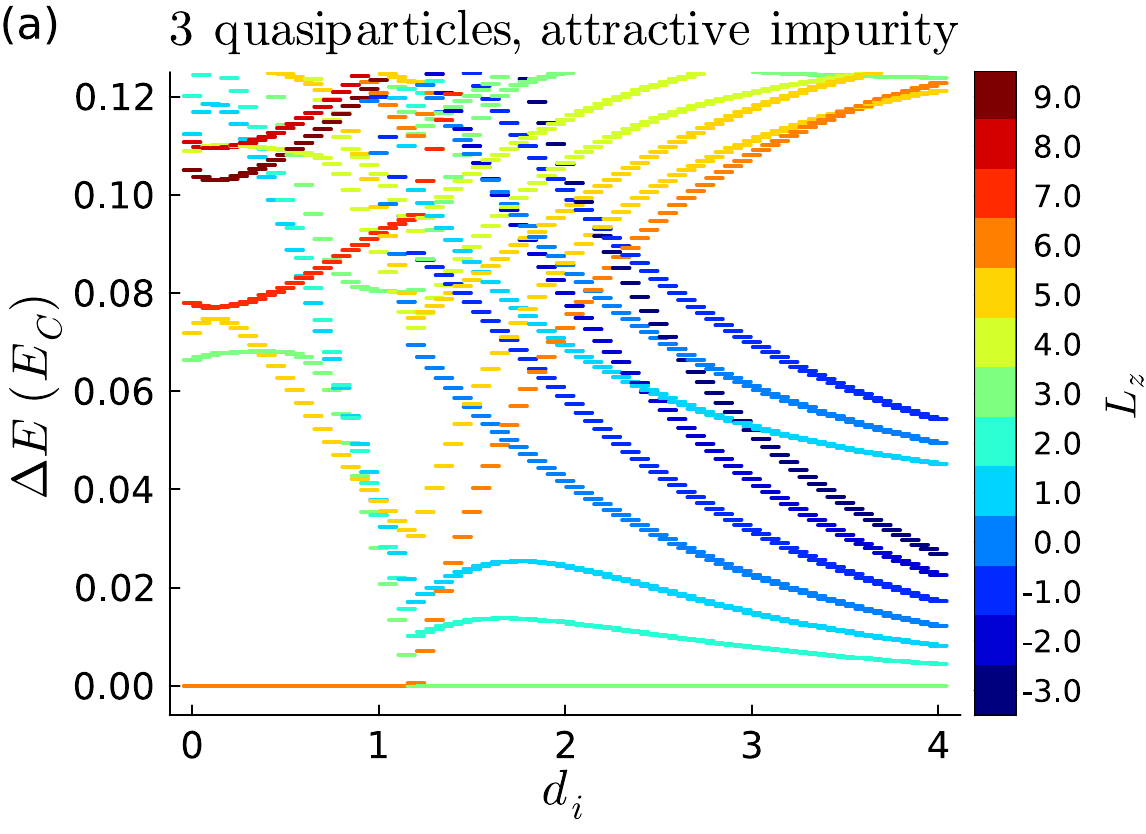}
\includegraphics[width=0.32\textwidth]{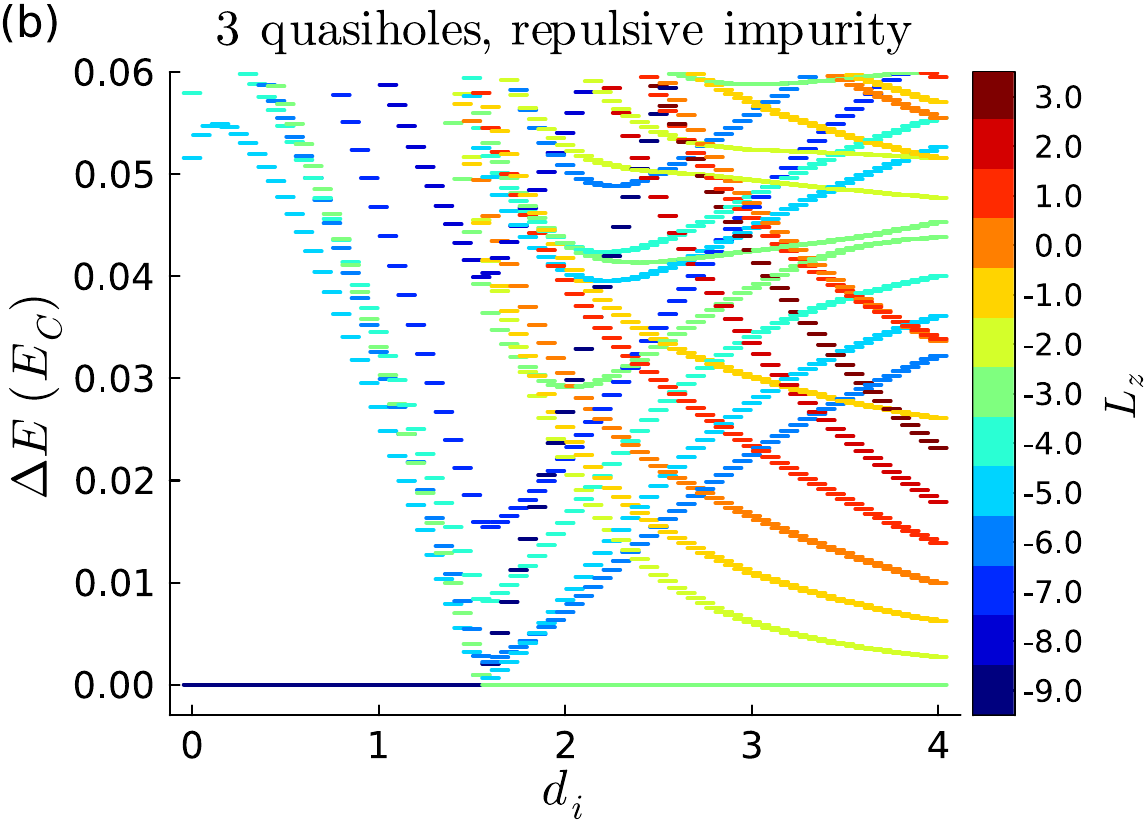}
\includegraphics[width=0.32\textwidth]{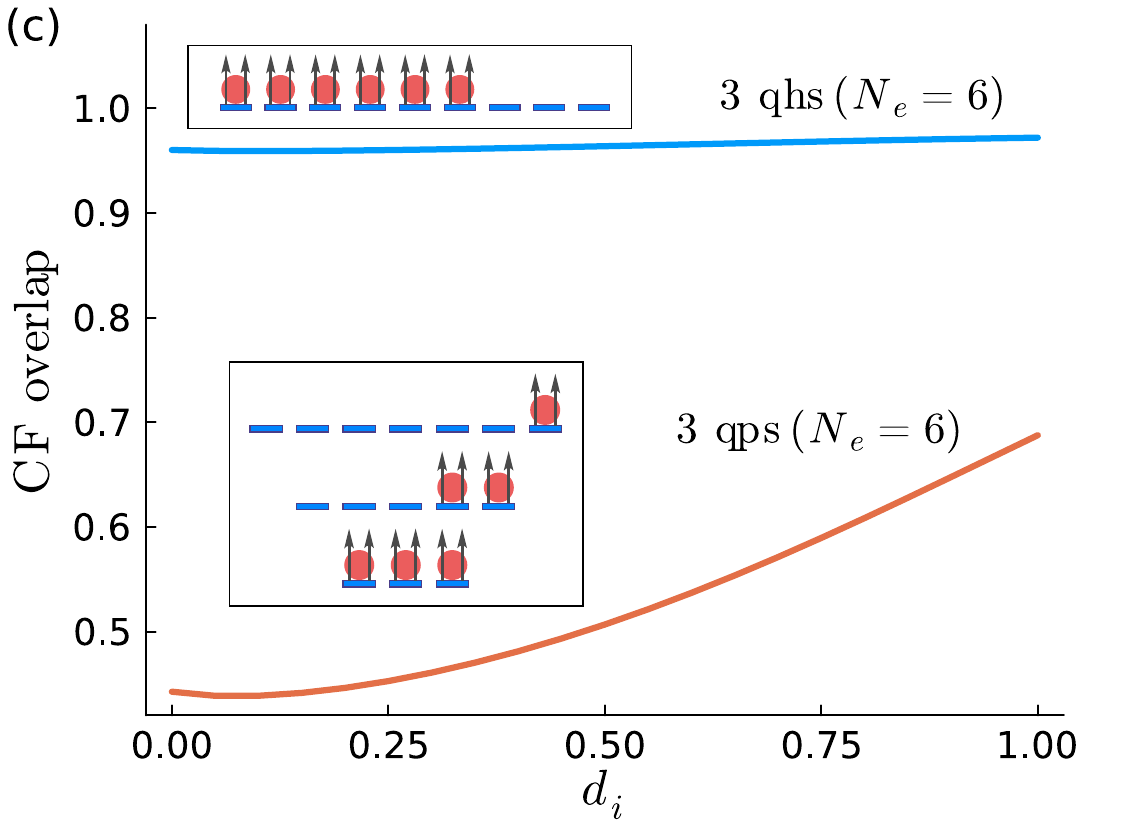}
\caption{
Energy spectrum of the $\nu=1/3$ state relative to the ground state, $\Delta E = E-E_0$, as a function of impurity distance $d_i$. Panel (a) shows the spectrum for \((N_e,2Q)=(6,12)\) with an attractive impurity (three additional quasiparticles), while panel (b) shows the spectrum for \((N_e,2Q)=(6,18)\) with a repulsive impurity (three additional quasiholes). The tip potential is placed directly above the impurity, while the gradient potential is not included. Both cases show a transition between the weak and strong impurity regimes, where the angular momentum of the ground state and the structure of the low-lying spectrum changes. (c) The squared overlap between the ground state and trial CF wave functions (insets) in the impurity distance range $0 \leq d_i \leq 1$. While the 3-quasihole ground state maintains very high overlap with the trial wave function, which shows little change with $d_i$, the surface impurity has a much stronger effect on the 3-quasiparticle state with less than 50\% overlap in the $d_i=0$ limit. The strong-impurity regime is therefore qualitatively different from the scenario considered in Ref.~\cite{Papic18}.}
\label{fig:distance_sweep}
\end{figure}

The IQH results are qualitatively independent of impurity distance $d_i$ from the sample. By contrast, Figures~\ref{fig:distance_sweep}(a)-(b) show that, as the impurity is brought towards the surface of the sample, the FQH state at $\nu=1/3$ exhibits a  spectral transition between two regimes that we dub the weak- and  strong-impurity regimes. This transition occurs for either three quasiparticles (a) or three quasiholes (b) present in the system, where we assume the impurity potential is attractive in the former case and repulsive in the latter. 

The weak-impurity case was studied in Ref.~\cite{Papic18}, where it was shown that it is well-described by composite fermion (CF) theory~\cite{Jain07} or exclusion rules based on Jack polynomials~\cite{Bernevig08}. However, upon reducing $d_i$ below $1-1.5\,\ell_B$, we find that the spectrum qualitatively reorganizes. Moreover, in the 3-quasiparticle case with strong impurity, the overlap of low-lying states with model wave functions based on CF theory drops significantly [Figure~\ref{fig:distance_sweep}(c)], suggesting a qualitatively different physics in the impurity-dominated regime in which our experiments take place. In the remainder we focus our attention to the 3-quasiparticle case that is relevant to experimental observations in the main text. 

\begin{figure}[!btp]
\centering
\includegraphics[width=0.32\textwidth]{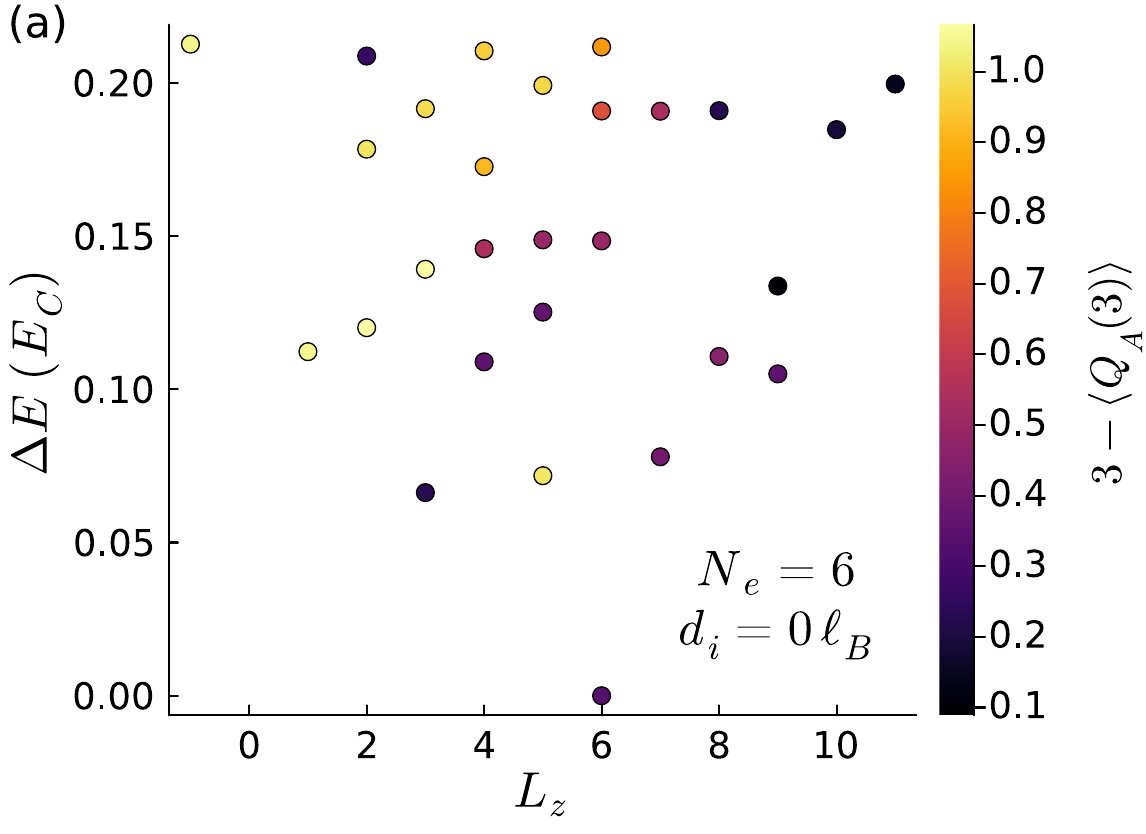}
\includegraphics[width=0.32\textwidth]{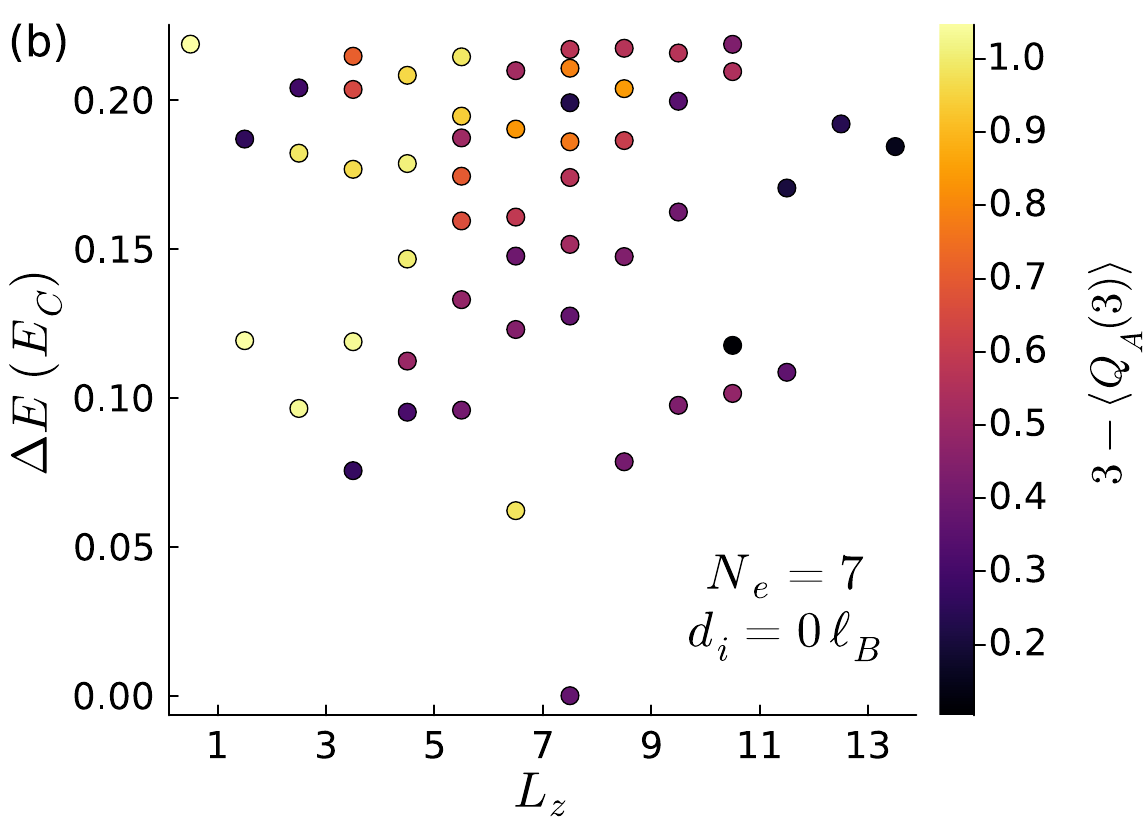}
\includegraphics[width=0.32\textwidth]{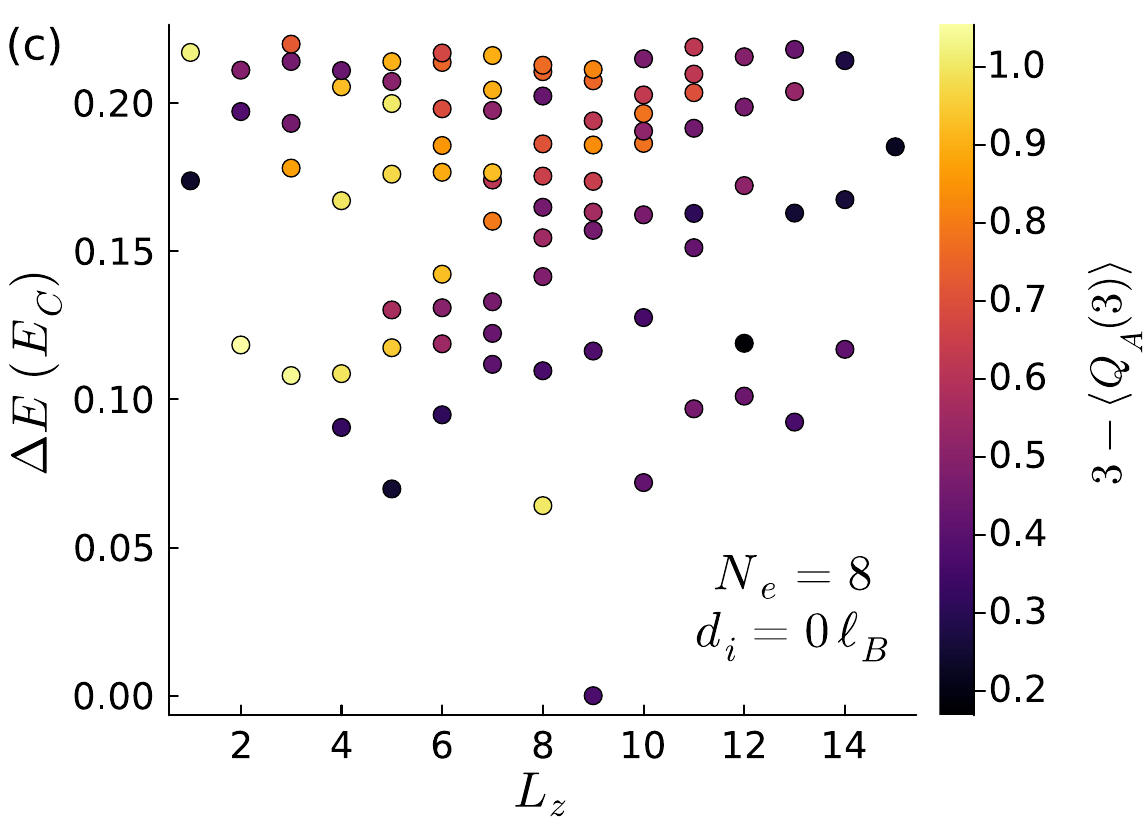}
\includegraphics[width=0.32\textwidth]{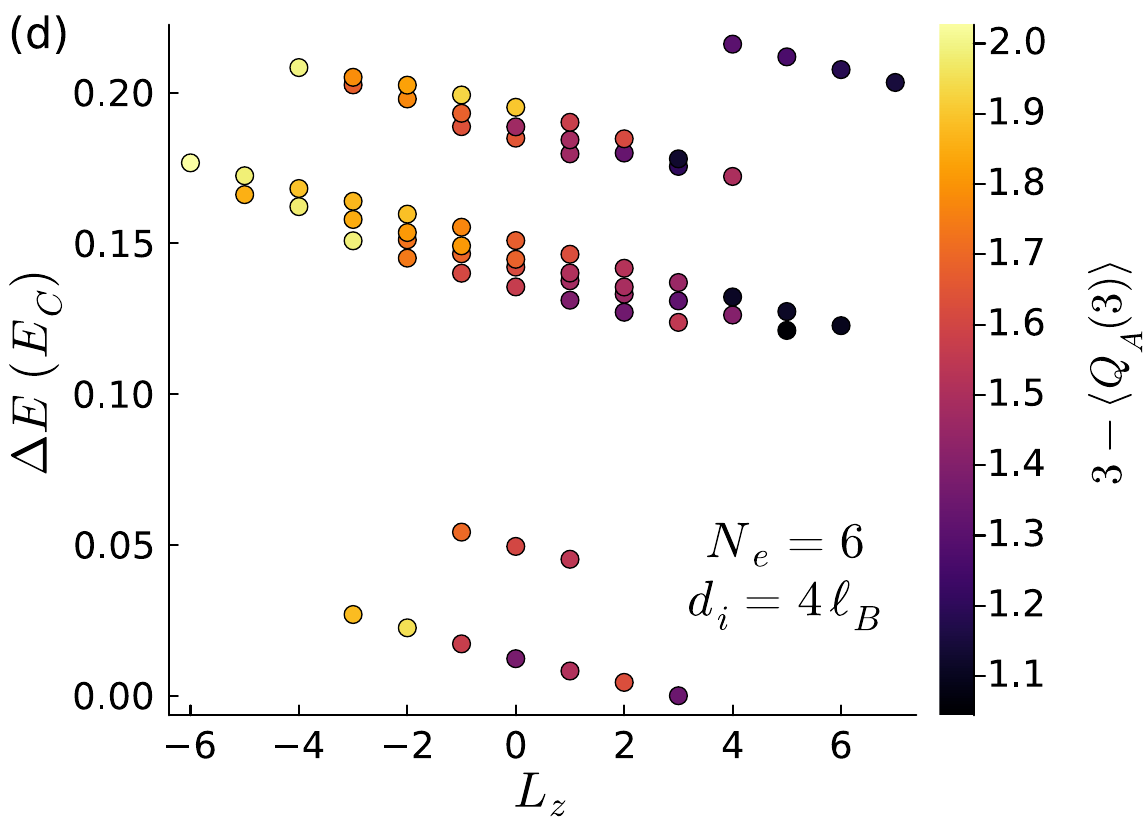}
\includegraphics[width=0.32\textwidth]{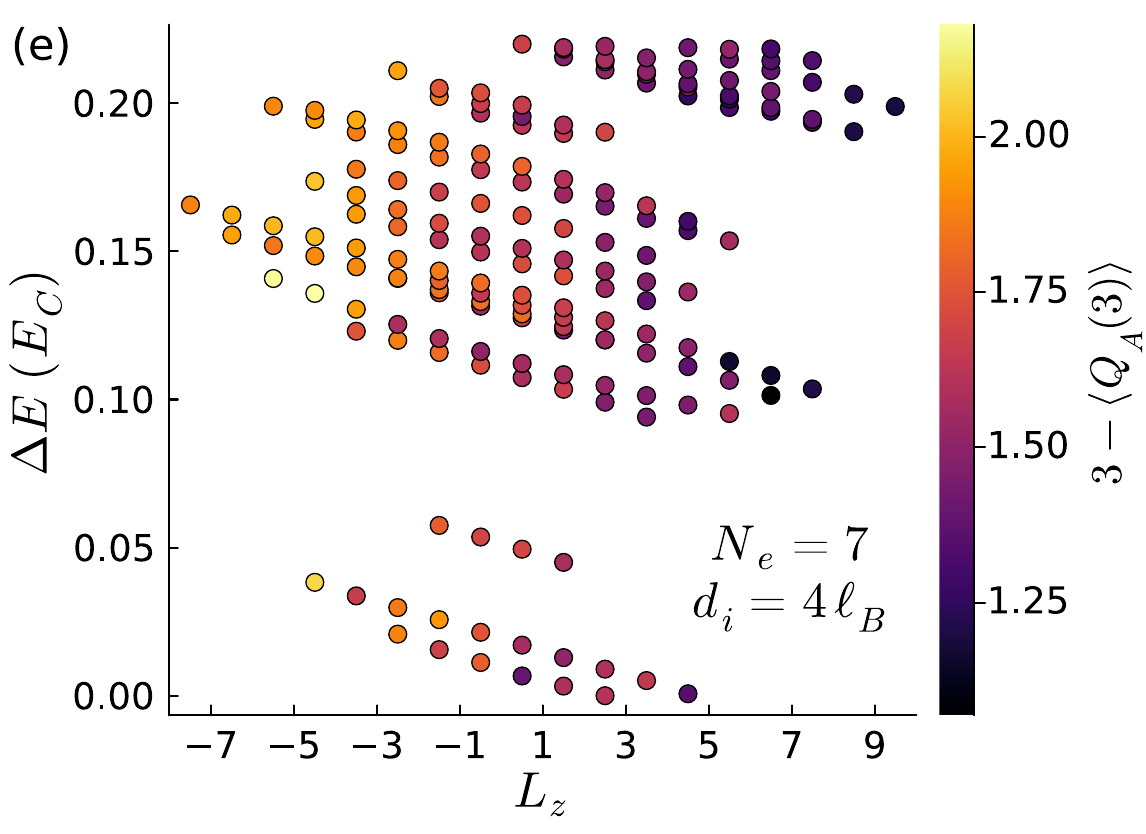}
\includegraphics[width=0.32\textwidth]{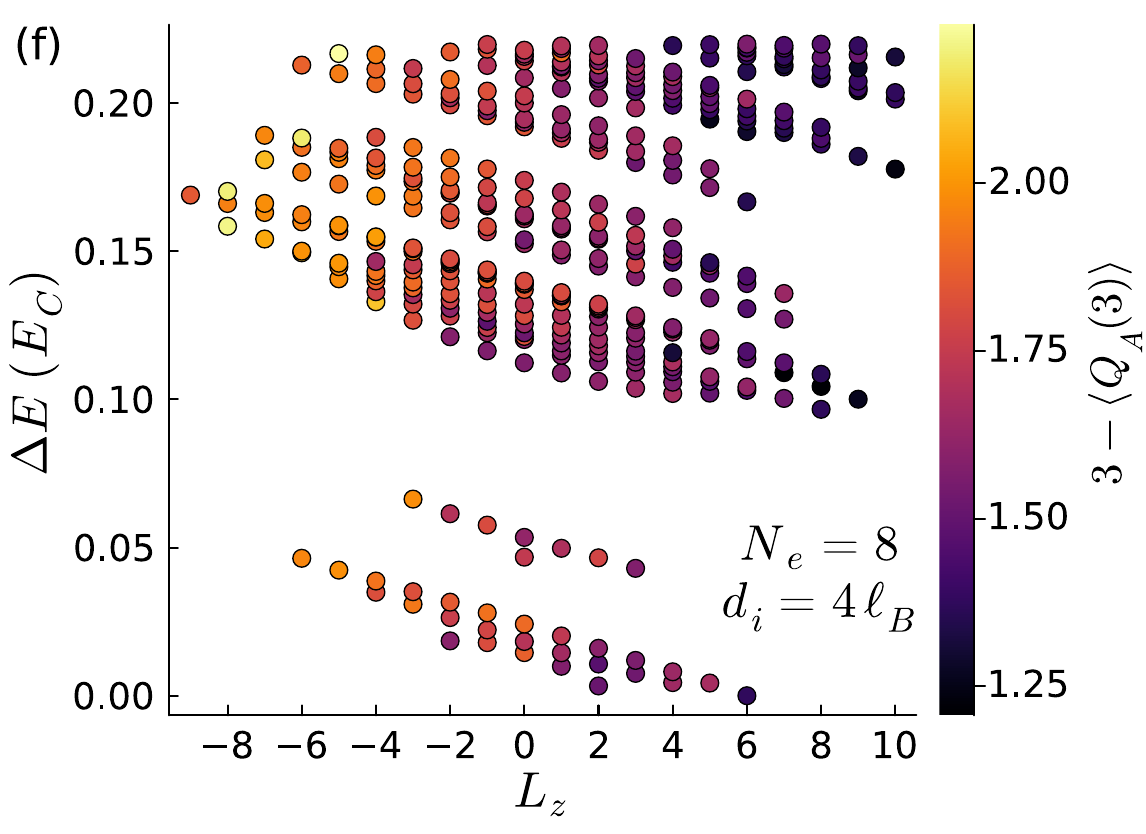}
\caption{Energy spectra resolved by \(L_z\) value for the  attractive impurity and three quasiparticles, for the system sizes $(N_e,2Q) = (6,12), (7,15), (8,18)$. Energies are plotted relative to the ground state, and the color bar shows the total charge deficit, $3 - \langle Q_A(3)\rangle$, on the three orbitals closest to the impurity. Top panels (a)-(c) are for the strong-impurity case $d_i=0$, bottom panels (d)-(f) are for the weak-impurity case $d_i=4\,\ell_B$.
}
\label{fig:qp_lz_spectra}
\end{figure}

The representative energy spectra at $d_i=0$ (strong-impurity regime) and $d_i=4\,\ell_B$ (weak-impurity regime) are shown in \Cref{fig:qp_lz_spectra} for the 3-quasiparticle case. 
In these plots, we include the impurity potential and place the tip on the impurity axis, but there is no gradient potential, hence the many-body angular momentum, $L_z = \sum_m m \hat n_m$, is  exactly conserved. The spectra in the two impurity regimes have qualitatively different structures: while the weak-impurity spectrum (considered in Ref.~\cite{Papic18}) is organized in approximately degenerate $L^2$ multiplets, the current strong impurity case lacks an apparent multiplet structure and instead features a unique ground state separated by a gap, whose $L_z$ value changes with $N_e$.  

%\begin{figure}[tbp]
%\centering
%\includegraphics[width=0.8\textwidth]{final_model_qh_fixed_distance_lz_spectra_di0_di4.pdf}
%\caption{
%Energy spectra resolved by \(L_z\) value for the  repulsive impurity and three quasiholes. We do not include the tip or gradient potentials, hence $L_z$ is conserved.   The left column uses \(d_i=0\), the right column uses
%\(d_i=4\ell_B\), and rows increase through
%\((N_e,2Q)=(4,12),(5,15),(6,18),(7,21)\).  Energies are plotted relative to the ground state at each $d_i$, \(E-E_0(d_i)\), and the color bar shows the total charge \(\langle Q_A(3)\rangle\) on the three orbitals closest to the impurity.
%}
%\label{fig:final-model-qh-fixed-distance-lz-spectra}
%\end{figure}

%\begin{figure}[tbp]
%\centering
%\includegraphics[width=0.8\textwidth]%{final_model_qp_fixed_distance_lz_spectra_di0_di4.pdf}
%\caption{
%Energy spectra resolved by \(L_z\) value for the  attractive impurity and three quasiparticles. We do not include the tip or gradient potentials, hence $L_z$ is conserved.  The left column uses \(d_i=0\), the right %column uses
%\(d_i=4\ell_B\), and rows increase through
%\((N_e,2Q)=(6,12),(7,15),(8,18),(9,21)\). Energies are plotted relative to the ground state at each $d_i$, \(E-E_0(d_i)\), and the color bar shows the total charge deficit, \(3-\langle Q_A(3)\rangle\), on the three orbitals %closest to the impurity.
%}
%\label{fig:final-model-qp-fixed-distance-lz-spectra}
%\end{figure}

%\begin{figure}[tbp]
%\centering
%\includegraphics[width=0.32\textwidth]{figs_final/spectrum_3qh_n_4_di_0.0.pdf}
%\includegraphics[width=0.32\textwidth]{figs_final/spectrum_3qh_n_5_di_0.0.pdf}
%\includegraphics[width=0.32\textwidth]{figs_final/spectrum_3qh_n_6_di_0.0.pdf}
%\includegraphics[width=0.32\textwidth]{figs_final/spectrum_3qh_n_4_di_4.0.pdf}
%\includegraphics[width=0.32\textwidth]{figs_final/spectrum_3qh_n_5_di_4.0.pdf}
%\includegraphics[width=0.32\textwidth]{figs_final/spectrum_3qh_n_6_di_4.0.pdf}
%\caption{Energy spectra resolved by \(L_z\) value for the  repulsive impurity and three quasiholes, for the system sizes $(N_e,2Q) = (4,12), (5,15), (6,18)$. Energies are plotted relative to the ground state, and the color bar shows the total charge, $\langle Q_A(3)\rangle$, on the three orbitals closest to the impurity. Top panels are for the strong-impurity case $d_i=0$, bottom panels are for the weak-impurity case $d_i=4$.
%}
%\label{fig:qh_lz_spectra}
%\end{figure}

We can further characterize the states according to the charge in a subsystem $A$, comprising some number of $M$ orbitals labelled from the impurity location at the north pole $m=Q$,
\begin{equation}\label{eq:QA}
    Q_A(M) = \sum_{m=0}^{M-1} n_{Q-m}.
\end{equation}
We will primarily be interested in the $M=3$ case. For the 3-quasiparticle case in \Cref{fig:qp_lz_spectra} we plot the charge deficit, $3-\langle Q_A\rangle$.  While in the 3-quasihole case, the ground state was previously shown to fully screen the impurity with all three $e/3$ quasiholes pinned to the pole~\cite{Wagner26}, we see this is not strictly true in the 3-quasiparticle case. This is consistent with the fact that FQH quasiparticles are more extended objects with a non-trivial internal structure compared to the quasiholes~\cite{Bernevig08a,Yang2014}. The interplay of this internal quasiparticle lengthscale with the decay of the impurity potential gives rise to a more complex ground state in the 3-quasiparticle case that does not have a simple representation, e.g., in terms of a variational wave function based on a single Jack polynomial root configuration.  

Similarly, the low-lying excited states in the strong-impurity case cannot be sharply distinguished by the amount of charge trapped around the impurity. As seen in \Cref{fig:qp_lz_spectra}(a)-(c), the low-lying states have similar values of $\langle Q_A(3)\rangle$ compared to the ground state, with the exception of a single state at momentum $L_z^0-1$, where $L_z^0$ denotes the ground state angular momentum. However, as we will demonstrate below, this special state with less trapped charge is not relevant for LDOS splitting as dipolar anisotropy is switched on.  

Once again, we emphasize that the 3-quasiparticle spectra in \Cref{fig:qp_lz_spectra} are markedly different from the 3-quasihole spectra (with repulsive impurity). The latter were recently studied in Ref.~\cite{Wagner26} where it was shown that the the low-lying excited states can indeed be distinguished by $\langle Q_A(M)\rangle$. For example, the first excited band of states consists of 2 quasiholes pinned at the impurity and the third quasihole displaced by some distance away (with the maximum allowed distance bounded by the sphere radius). Due to the simple structure of FQH quasihole excitations, this picture remarkably holds even in the impurity-dominated regime where the ground state and low-lying excited states can still be accurately identified with a single root configuration~\cite{Rezayi85,Wagner26}. Due to the complexity of FQH quasiparticles mentioned above, this simple picture no longer holds in our case, as also indicated by the low overlap of the CF wave function with the ground state in the strong-impurity regime in \Cref{fig:distance_sweep}(c).

\subsubsection{The breaking of rotational symmetry}

\begin{figure}[!tbp]
\centering
\includegraphics[width=0.4\textwidth]{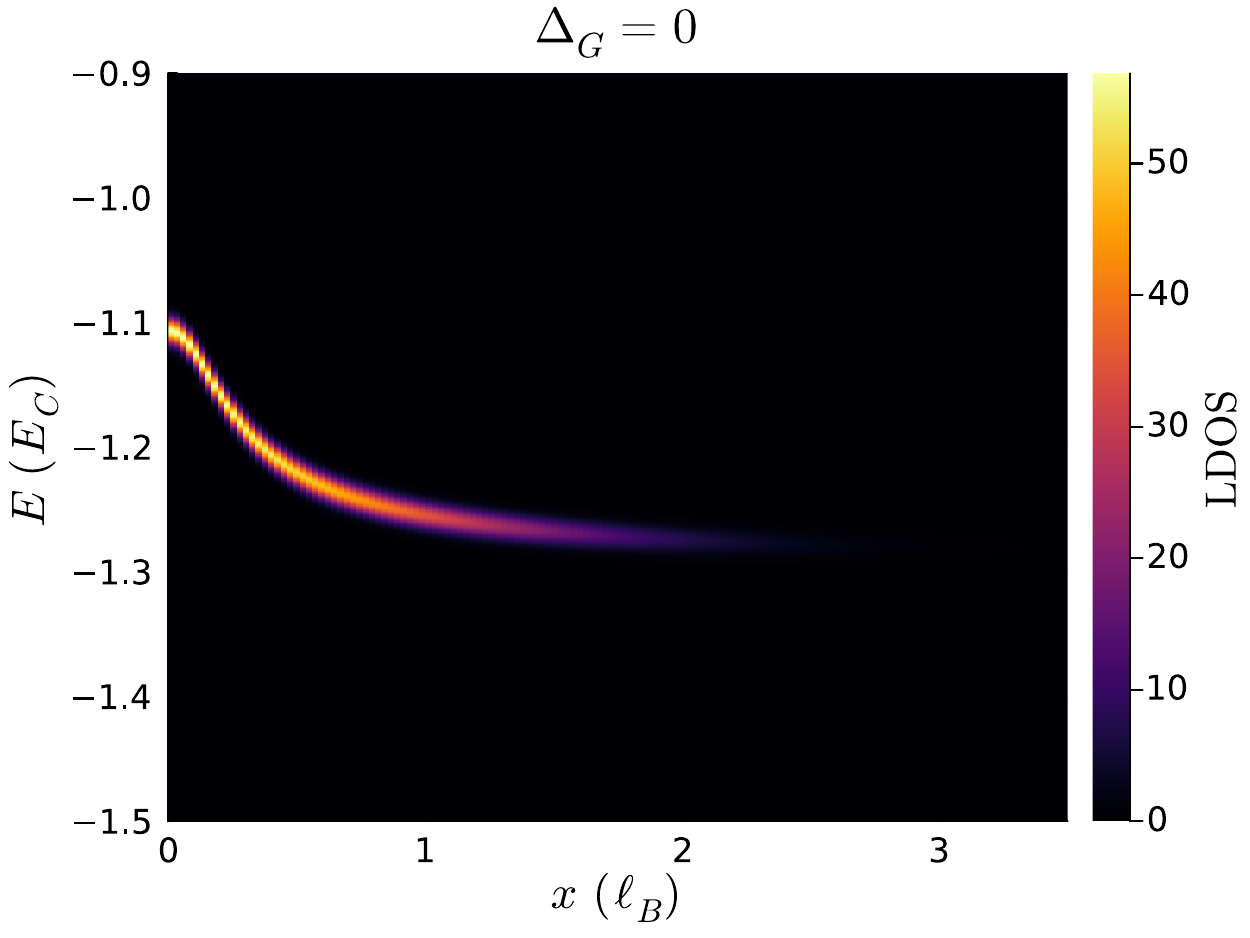}
\includegraphics[width=0.4\textwidth]{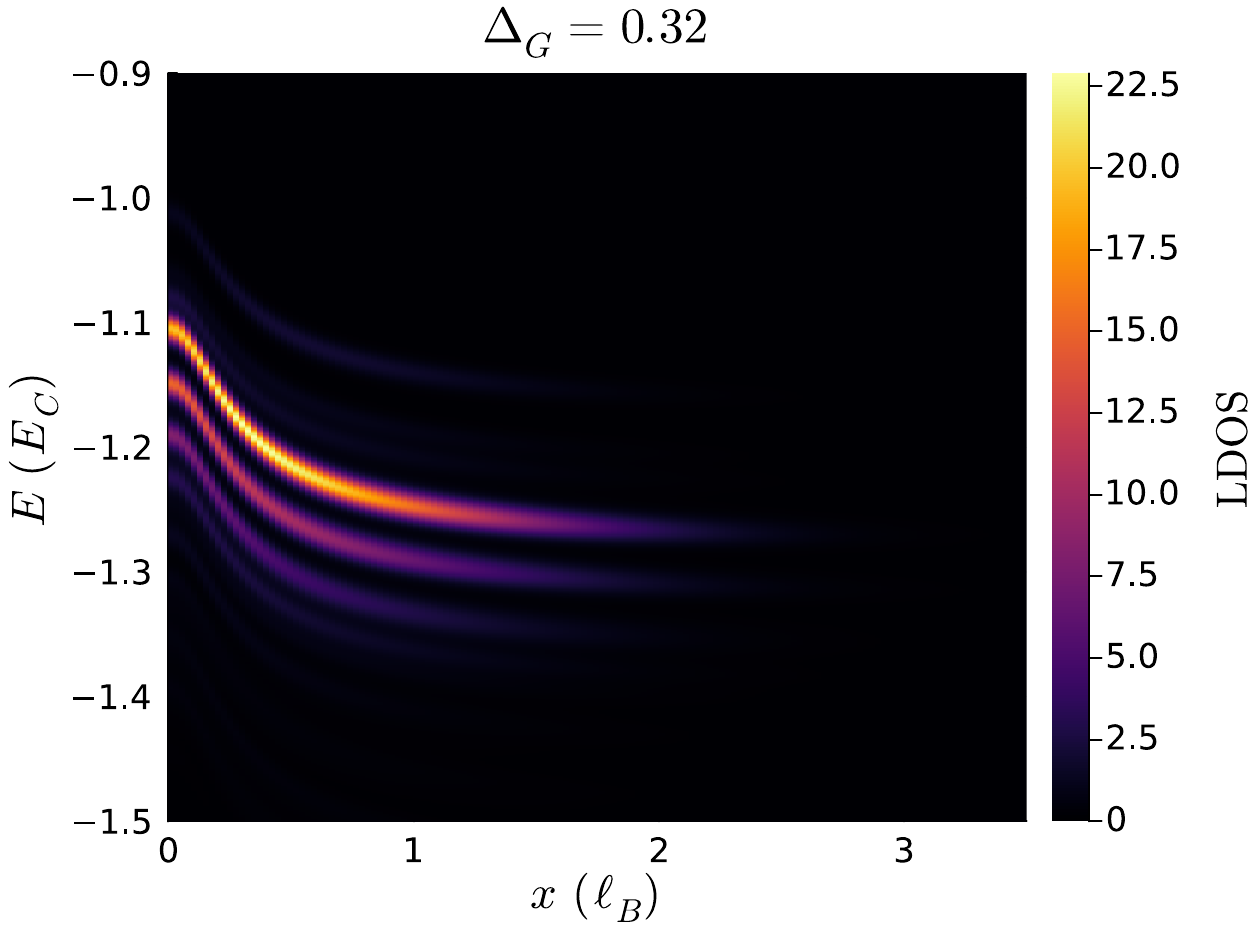}
\caption{
Spatial LDOS maps for the electron removal from a  \(N_e=6\), \(2Q=12\) ground state with strong impurity (the 3-quasiparticle case) at \(\Delta_G=0\) (left panel) and
\(\Delta_G=0.32\) (right panel). The tip is fixed at \(d_t=0.2\ell_B\).  The plotted
\(E/E_C\) are measured relative to the chemical potential. In both cases, the LDOS spectral weight is dominated by a single or a few bright peaks, although much weaker features are present even beyond the energy scale of the figures.
}
\label{fig:final-model-qp-ldos-fields-htip045}
\end{figure}

The spectra in \Cref{fig:qp_lz_spectra} include the impurity and tip effects (placed directly above the impurity), but no gradient potential, hence there is exact rotational symmetry about the $z$-axis. In this scenario, the LDOS spectrum for the electron removal from a 3-quasiparticle ground state in the strong-impurity regime shows a single peak, as confirmed by the numerics in the left panel of Figure~\ref{fig:final-model-qp-ldos-fields-htip045}. We have confirmed the existence of a single peak in the rotationally-symmetric case for larger sizes up to $N_e\leq 9$ (data not shown).

By contrast, setting $\Delta_G=0.32$, the LDOS peak splits into several ones with varying degrees of intensity, see the right panel of Figure~\ref{fig:final-model-qp-ldos-fields-htip045}. This shows that anisotropy  can produce a splitting of dominant LDOS peaks in the $\nu=1/3$ state. While Figure~\ref{fig:final-model-qp-ldos-fields-htip045} is for three flux quanta less than the vacuum ($2Q=3N_e-6$), the splitting is not unique to this case; we also observe it numerically with fewer flux quanta removed (e.g., at $2Q=3N_e-5$), as well as with the same numbers of added flux quanta and a repulsive impurity. 

\begin{figure}[!bth]
\centering
\includegraphics[width=0.4\textwidth]{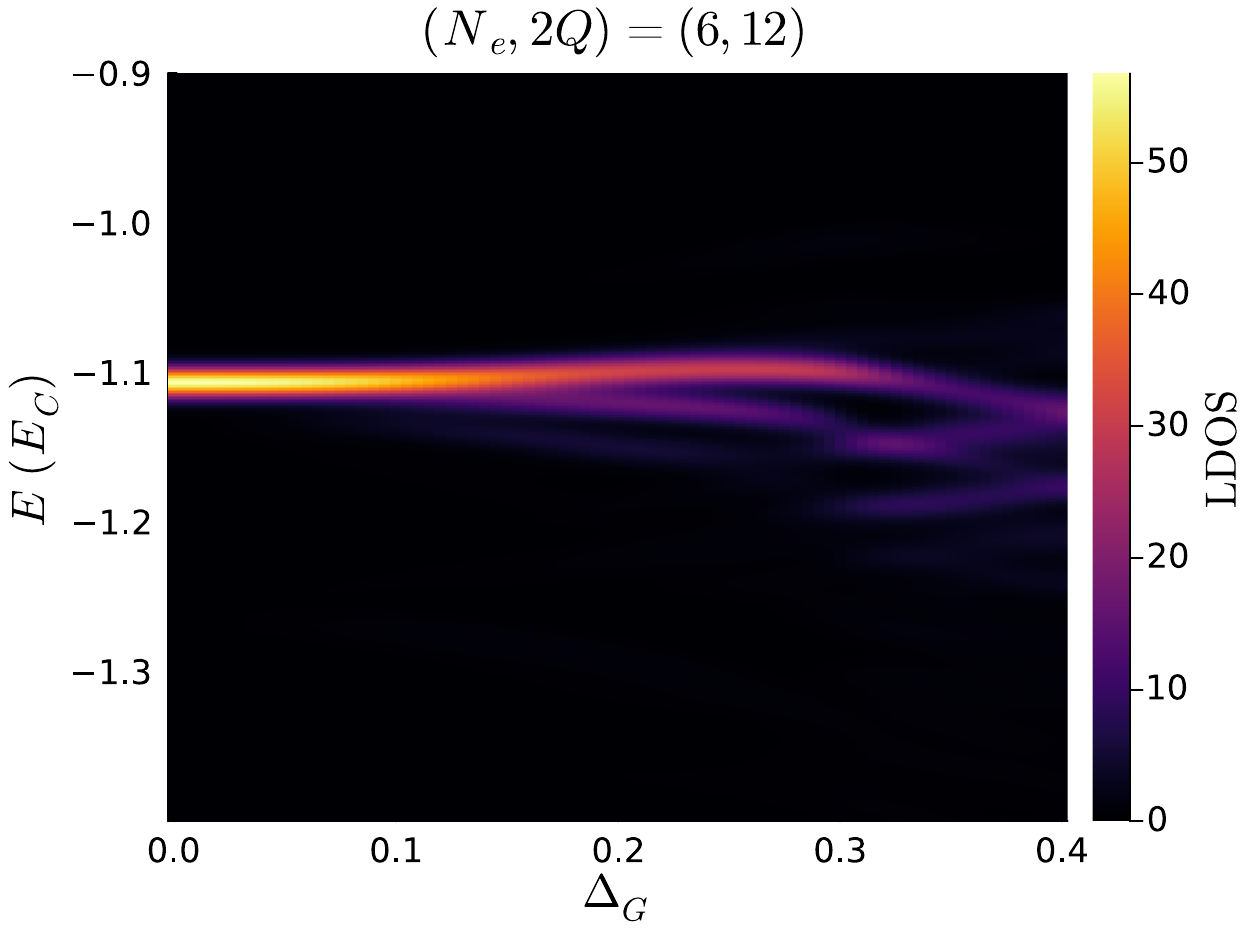}
\includegraphics[width=0.4\textwidth]{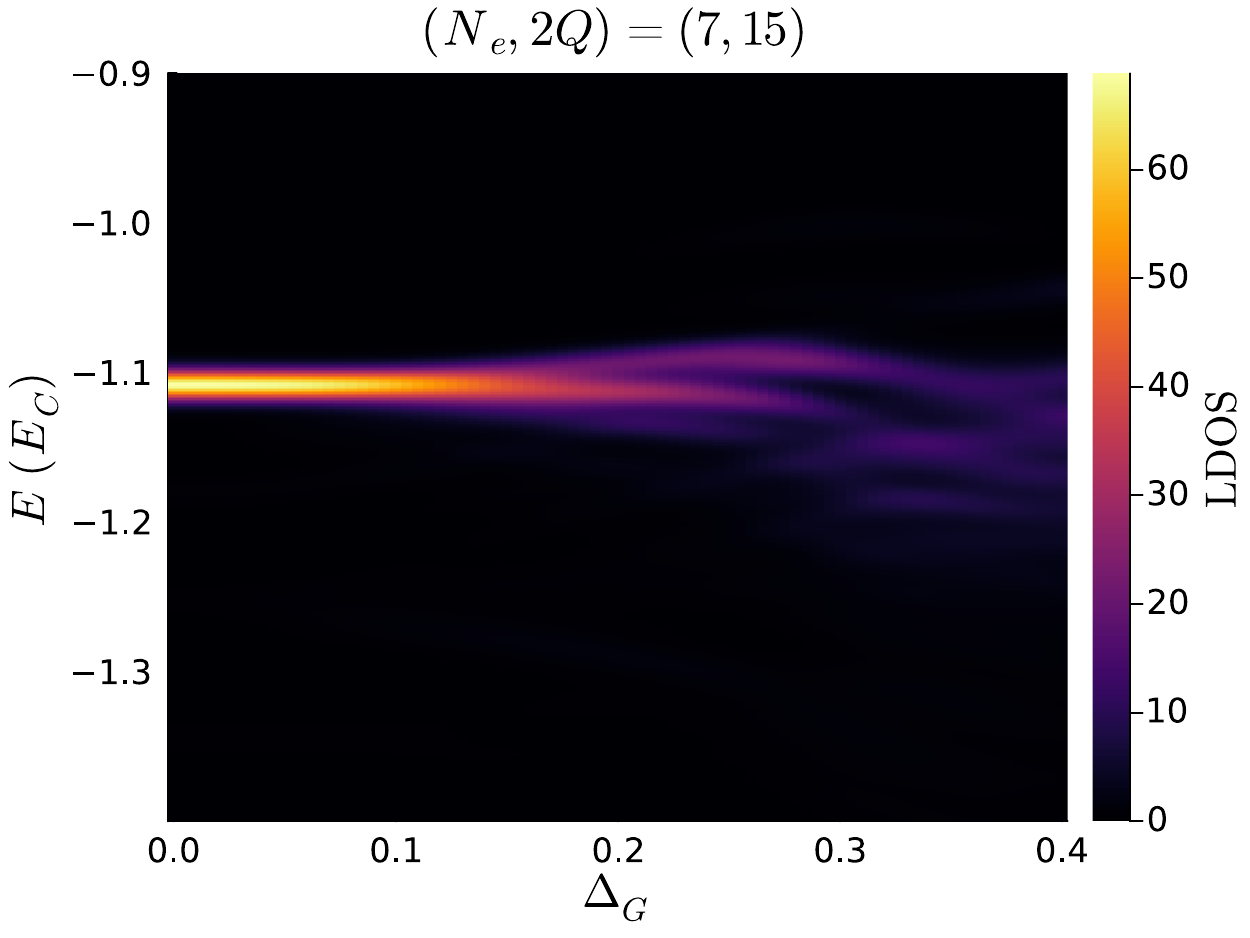}
\caption{
The evolution of the LDOS peak intensity at \(x=0\) as a function of \(\Delta_G\) for
\((N_e,2Q)=(6,12)\) and \((7,15)\).  Both panels use \(d_i=0\,\ell_B\),
\(d_g=7\,\ell_B\), \(d_t=0.2\,\ell_B\). As the anisotropy is increased, the main resonance splits into multiple peaks.}
\label{fig:final-model-tip-x0-ldos-delta-g}
\end{figure}

In Figure~\ref{fig:final-model-tip-x0-ldos-delta-g} we study more systematically the evolution of the bright LDOS peak slice at $x=0$ upon varying anisotropy $\Delta_G$. The qualitative behavior is consistent between the two system sizes, with the splitting occurring over a broad range of $\Delta_G\gtrsim 0.3$. In the experimental setup, anisotropy is intrinsically present due to the impurity and the image charge of the tip. Hence, this simulation captures the experimental phenomenology from the main text.

\subsubsection{The mechanism of LDOS splitting}

\begin{figure}[!htbp]
\centering
\includegraphics[width=0.4\textwidth]{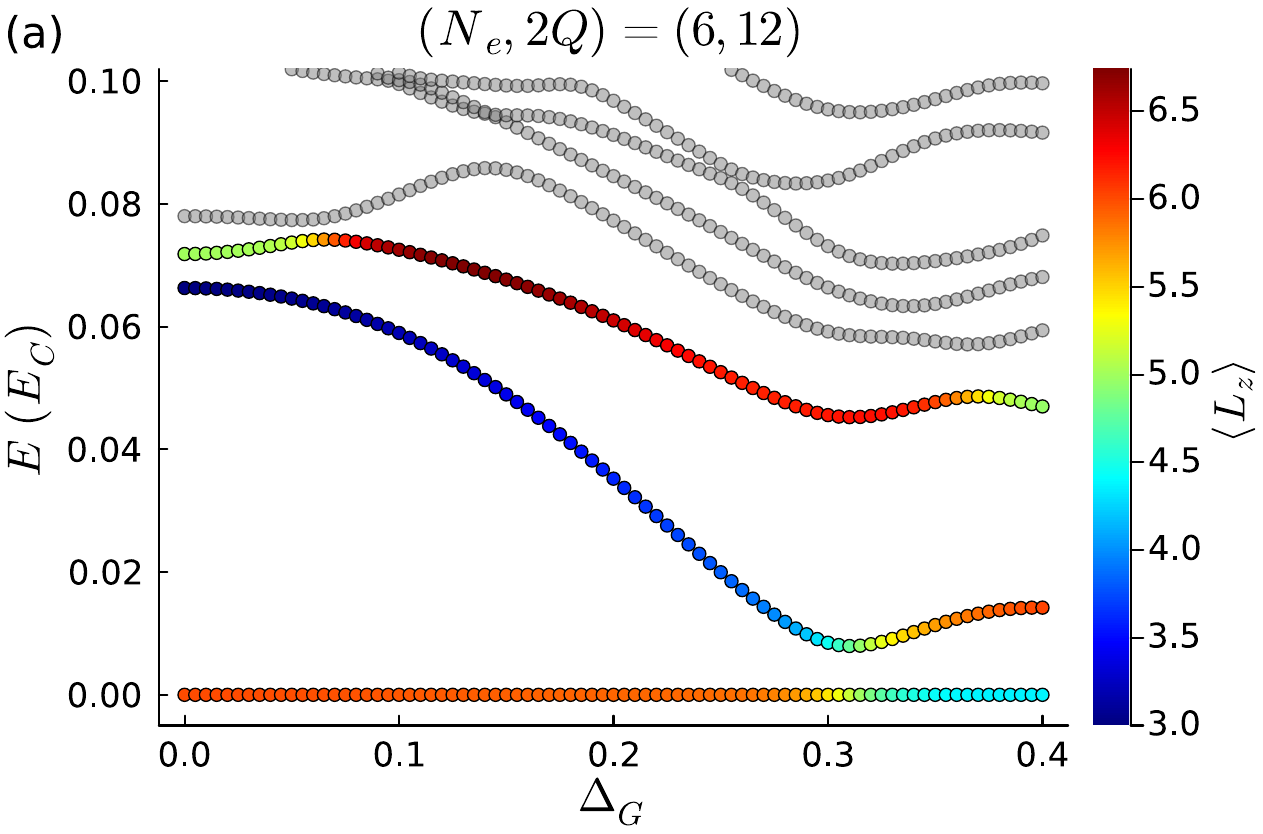}\hspace{0.05\textwidth}
\includegraphics[width=0.4\textwidth]{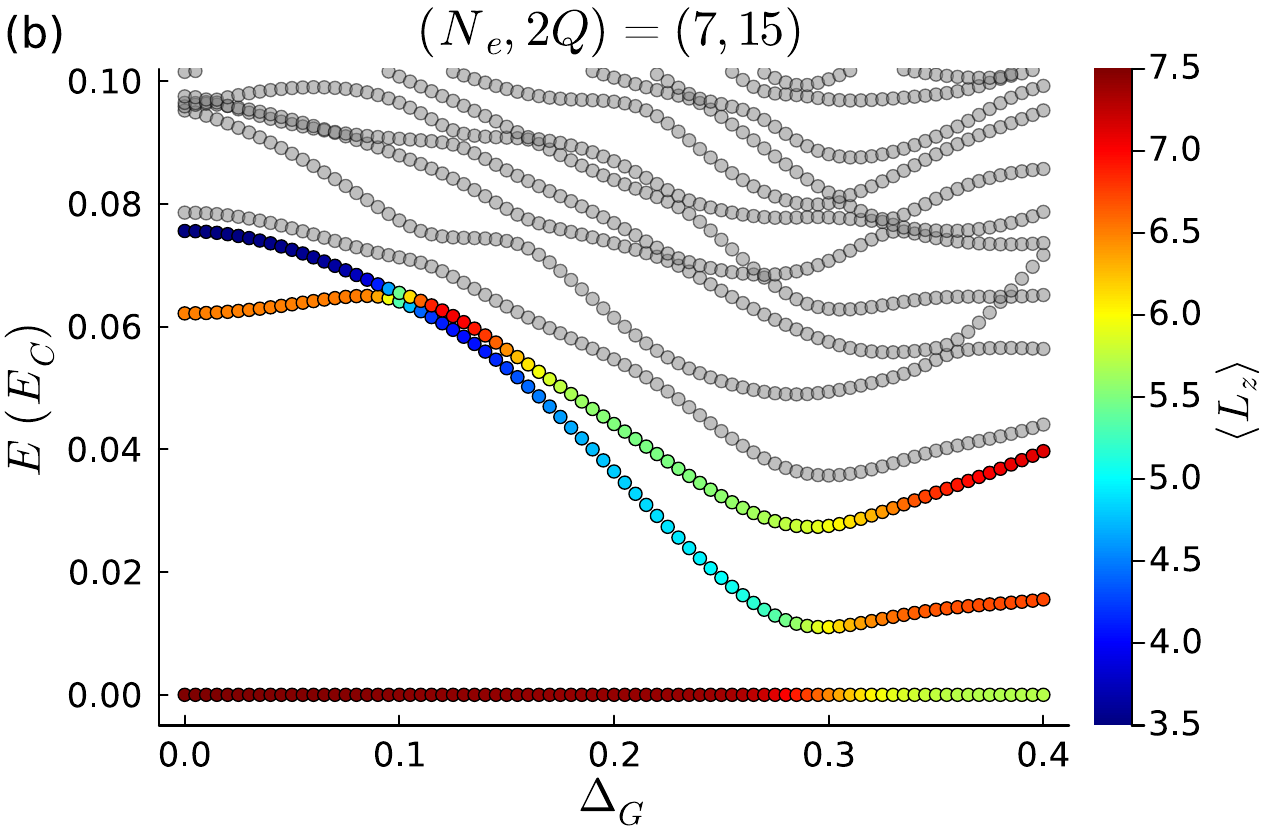}\\[1.0ex]
\includegraphics[width=0.4\textwidth]{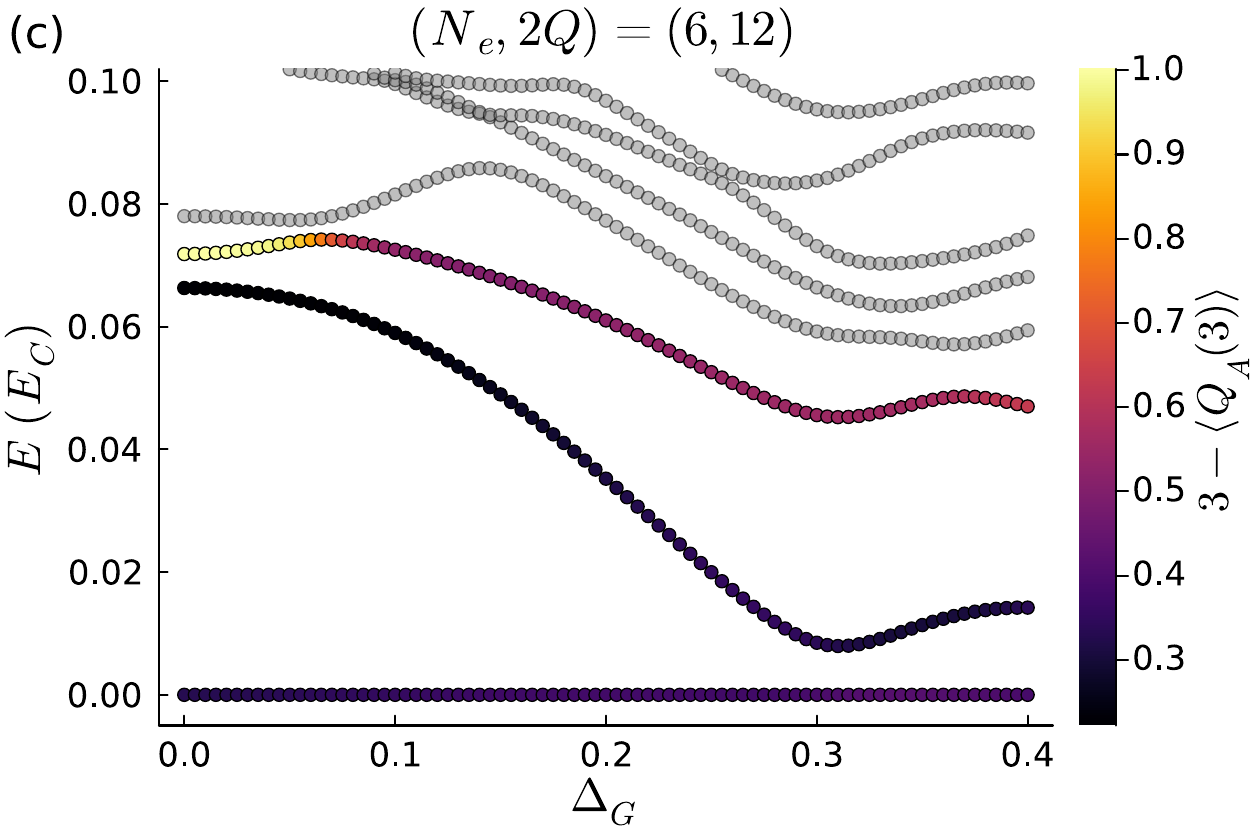}
\hspace{0.05\textwidth}
\includegraphics[width=0.4\textwidth]{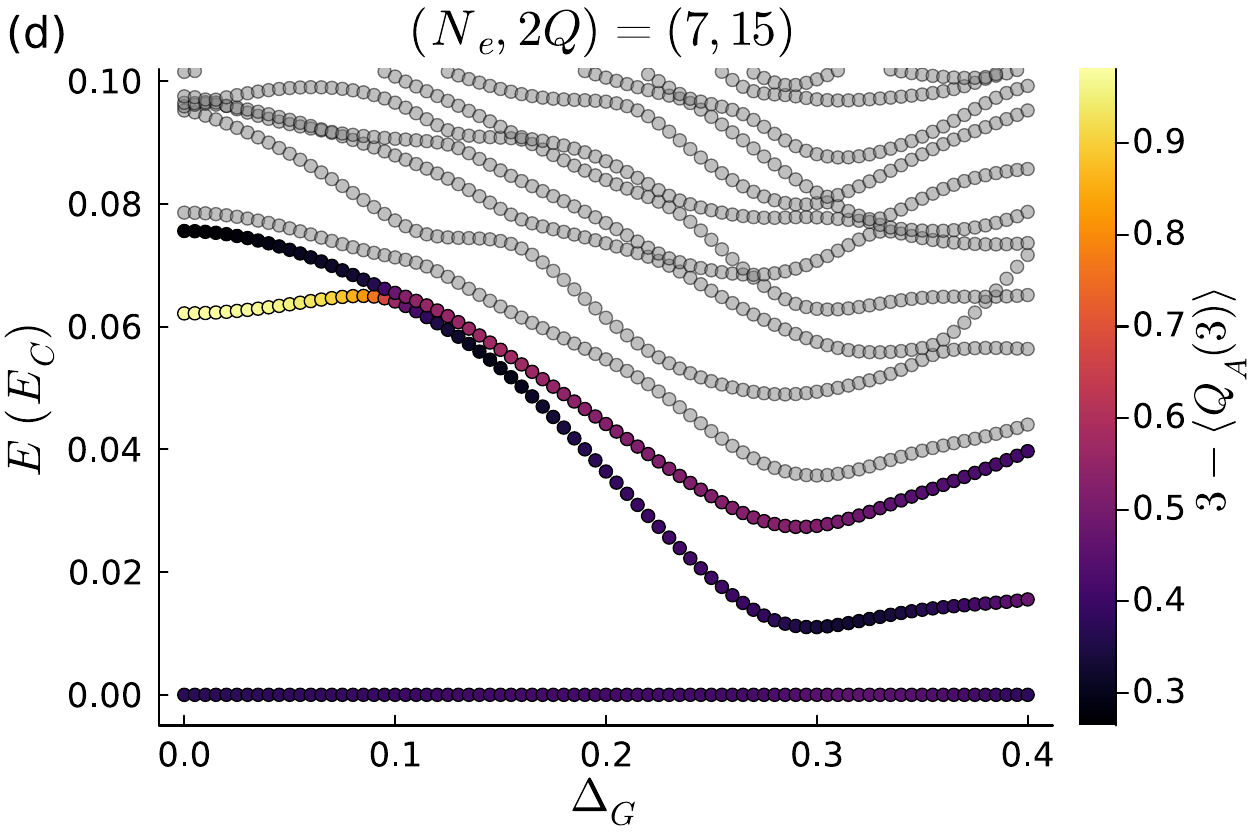}
\caption{
The evolution of energy spectra as a function of gradient $\Delta_G$ for the attractive impurity at $d_i=0\,\ell_B$ and \(3\)-quasiparticles present. Other parameters are $d_g=7\,\ell_B$, $d_t=0.2\,\ell_B$, \(\theta_{\mathrm{tip}}=0\). Panels (a),(c) show the spectrum at system size $(N_e,2Q)=(6,12)$, where in (a) the lowest three states are colored by their expected angular momentum $\langle L_z \rangle$, while in (c) the color is given by the average charge depletion in the first three orbitals $3 - \langle Q_A(3)\rangle$. Panels (b), (d) show the same data for the next larger system size $(N_e,2Q)=(7,15).$
}
\label{fig:final-model-qp-tip-gradient-focus-spectra}
\end{figure}

Finally, we study the impact of rotation-symmetry breaking fields on the energy spectra. While the final energy spectra (after the tunneling event) have a significant impact on the LDOS, the resonances above the impurity appear at high energies. As they are deep inside the continuum of excited states, their behaviour is difficult to track microscopically; therefore, we will focus on the initial spectra, before tunneling. For simplicity, we fix the impurity distance to $d_i=0$ and $d_t=0.2$, \(\theta_{\mathrm{tip}}=0\), while varying $\Delta_G$. The energy spectra are shown in Figure~\ref{fig:final-model-qp-tip-gradient-focus-spectra} for the 3-quasiparticle case. Each panel tracks the lowest three states in the spectrum and colors them according to their average $\langle L_z\rangle$ value (no longer exactly conserved) as well as the local charge deficit,  $3-\langle Q_A(3)\rangle$. 

Based on this data, we attribute the LDOS splitting to the avoided crossing  in the spectrum: compare the top left panel of Figure~\ref{fig:final-model-qp-tip-gradient-focus-spectra} with the LDOS map in Figure~\ref{fig:final-model-qp-ldos-fields-htip045}, which are for the same system size and choice of parameters. The LDOS splitting around $\Delta_G=0.3$ coincides with the avoided crossing between the ground state with $\langle L_z\rangle \approx 6$ and the first excited state with $\langle L_z\rangle \approx 3$, which mix and change their characters near the same value $\Delta_G=0.3$. A similar pattern is observed in the larger system size, $N_e=7$, where another avoided crossing occurs earlier on, at about $\Delta_G\sim 0.1$; however, since this involves the first two excited states, it does not visibly impact the LDOS in Figure~\ref{fig:final-model-qp-ldos-fields-htip045}, with the main splitting still occurring slightly below $\Delta_G=0.3$.

%\begin{figure}[!tbp]
%\centering
%\includegraphics[width=0.4\textwidth]{figs_final/lz_evolution_n_5_3qhs.pdf}\hspace{0.05\textwidth}
%\includegraphics[width=0.4\textwidth]{figs_final/lz_evolution_n_6_3qhs.pdf}\\[1.0ex]
%\includegraphics[width=0.4\textwidth]{figs_final/avg_charge_evolution_n_5_3qhs.pdf}
%\hspace{0.05\textwidth}
%\includegraphics[width=0.4\textwidth]{figs_final/avg_charge_evolution_n_6_3qhs.pdf}
%\caption{
%The evolution of energy spectra as a function of gradient $\Delta_G$ for the repulsive impurity at $d_i=0\,\ell_B$ and \(3\)-quasiholes present. Other parameters are $d_g=7\,\ell_B$, $d_t=0.2\,\ell_B$, \(\theta_{\mathrm{tip}}=0\). Panels (a),(c) show the spectrum at system size $(N_e,2Q)=(5,15)$, where in (a) the lowest three states are colored by their expected angular momentum $\langle L_z \rangle$, while in (c) the color is given by the average charge in the first three orbitals $\langle Q_A(3)\rangle$. Panels (b), (d) show the same data for the next system size $(N_e,2Q)=(6,18).$}
%\label{fig:final-model-qh-tip-gradient-focus-spectra}
%\end{figure}

\begin{figure}[!htbp]
\centering
\includegraphics[width=0.42\textwidth]{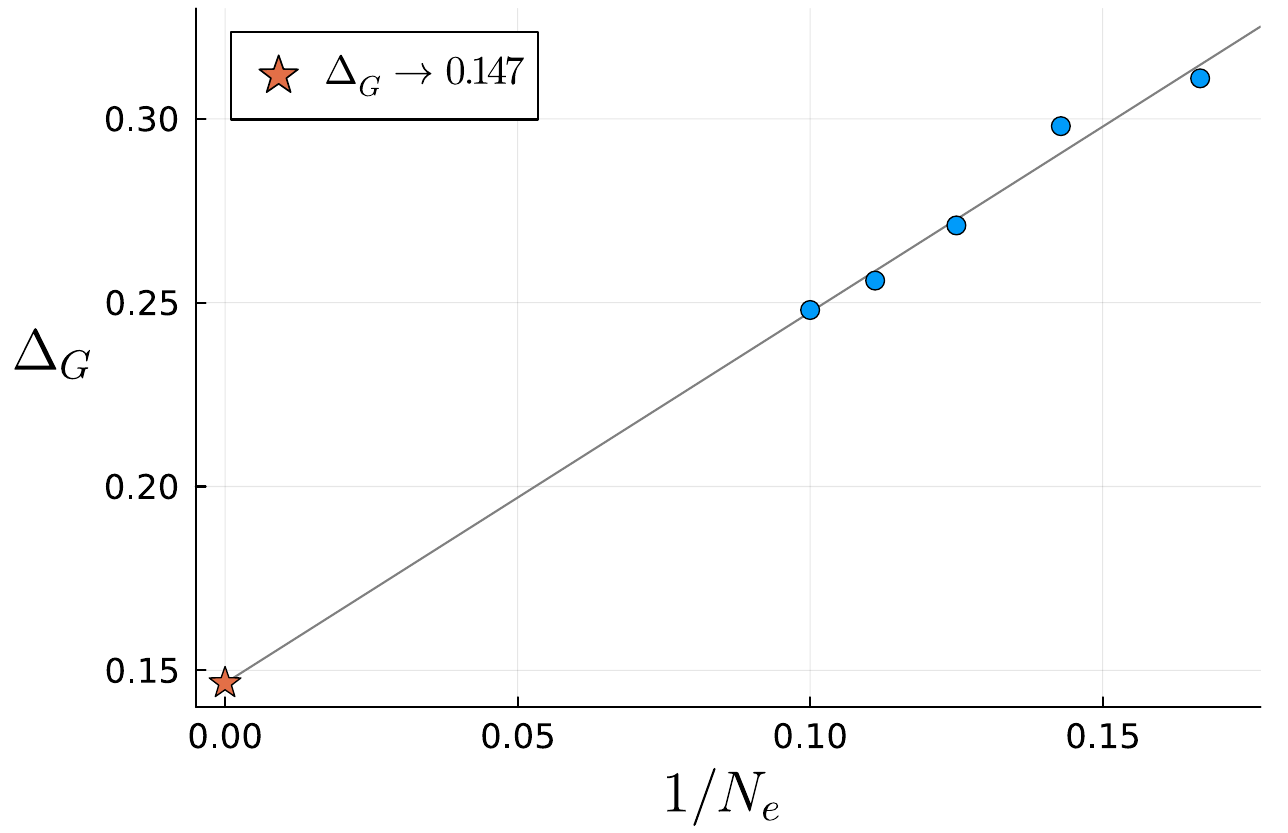}
\caption{
Finite-size linear extrapolation of \(\Delta_G\) at the point of the avoided crossing in the 3-quasiparticle case in Figure~\ref{fig:final-model-qp-tip-gradient-focus-spectra}. %In the thermodynamic limit, the value of $\Delta_G$ where the STM spectrum develops multiple peaks at the impurity roughly coincides with the value extracted from the $\nu=1$ real-space maps. 
}
\label{fig:delta_g_extrapolation}
\end{figure}

The finite-size dependence of $\Delta_G$ inducing LDOS splitting is presented in Figure~\ref{fig:delta_g_extrapolation}. Since producing LDOS maps is computationally demanding due to the spatial dependence of the tip potential, we turn to the spectrum at the fixed $d_t=0.2\,\ell_B$ and examine the point of avoided level crossing as a
shallow minimum, whose position shifts to smaller \(\Delta_G\) with increasing
system size.  Extracting the point of avoided crossing by a local quadratic
interpolation and fitting the resulting \(\Delta_G^\star\) values linearly in
\(1/N_e\) gives \(\Delta_G^\star(\infty)\approx 0.15\).

\begin{figure}[!b]
\centering
\includegraphics[width=0.32\textwidth]{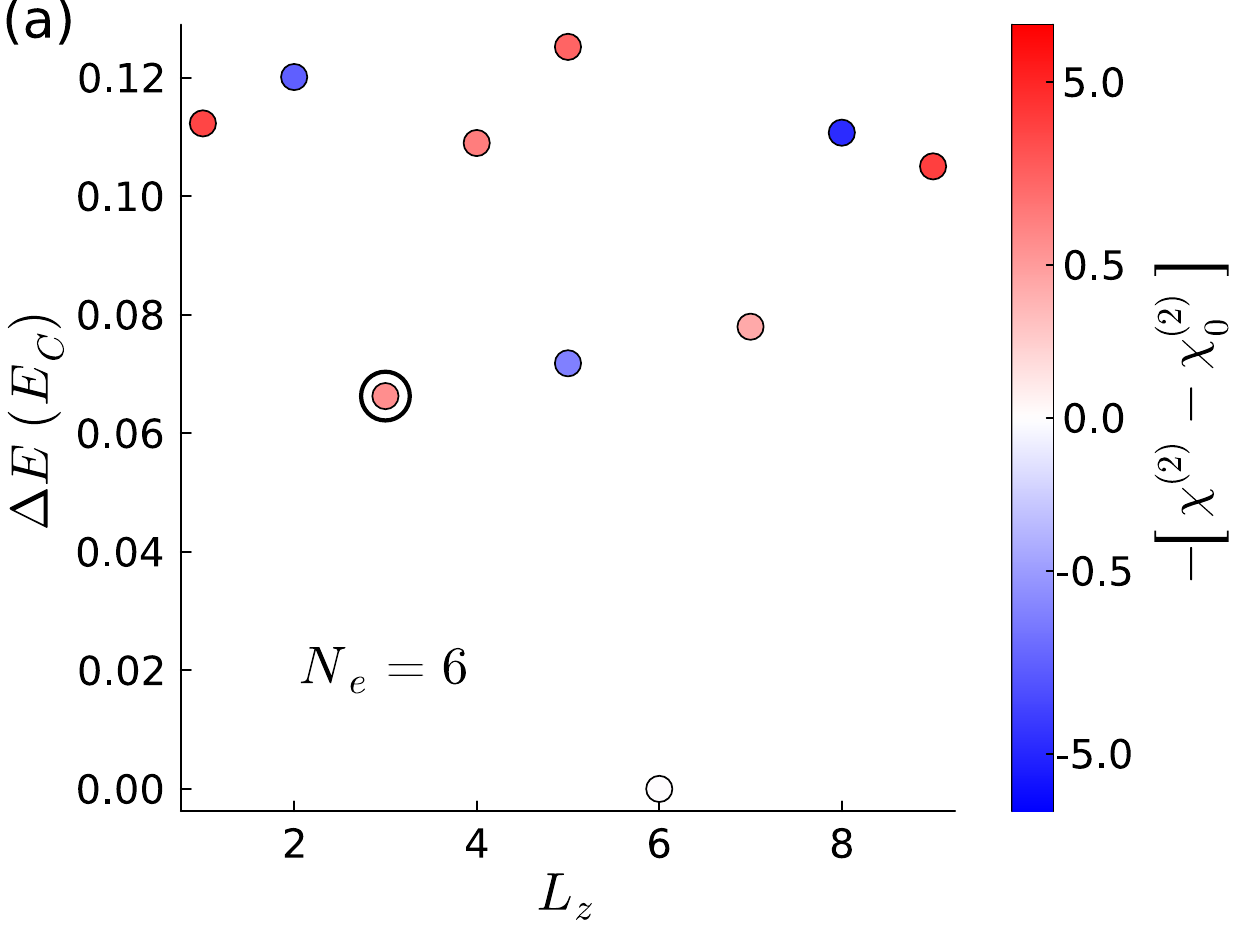}\hfill
\includegraphics[width=0.32\textwidth]{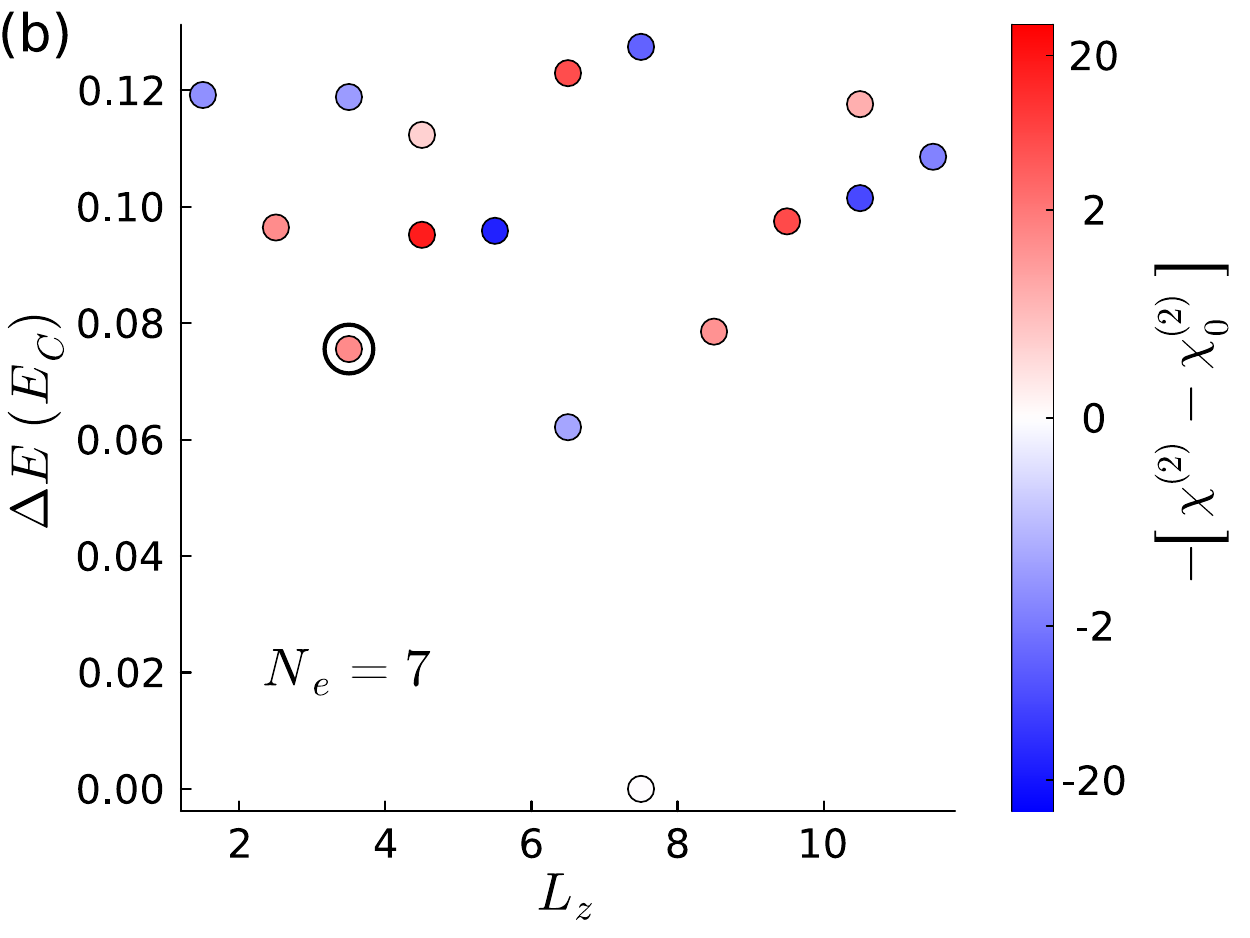}\hfill
\includegraphics[width=0.32\textwidth]{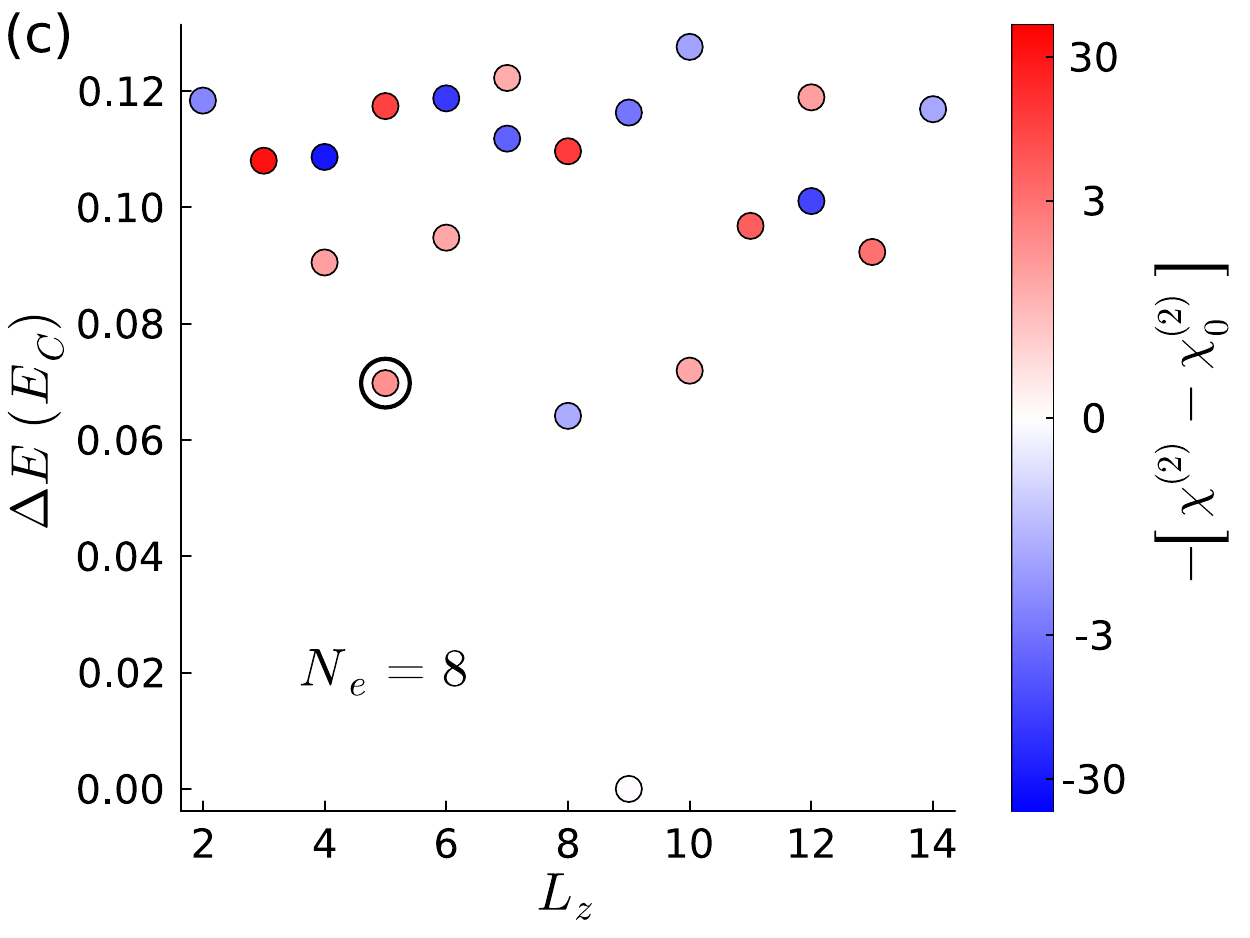}
\caption{
Susceptibility diagnostic, Eq.~(\ref{eq:susc}), for the
\(3\)-quasiparticle spectra at \(d_i=0\,\ell_B\), \(d_g=7\,\ell_B\),
\(d_t=0.2\,\ell_B\) and $\Delta_G=0$.  Panels (a)-(c) show
\((N_e,2Q)=(6,12),(7,15),(8,18)\).  The vertical axis is the excitation
energy at \(\Delta_G=0\), and the color shows the signed-log value of
\(-[\chi_a^{(2)}-\chi_0^{(2)}]\), so red states lower their gap relative to
the ground state under a small anisotropy field.  The black circles mark the
state identified to undergo the avoided-crossing  in
Figure~\ref{fig:final-model-qp-tip-gradient-focus-spectra}. 
}
\label{fig:final-model-qp-gradient-susceptibility}
\end{figure}

To further isolate the low-lying state most relevant for the avoided crossing in Figure~\ref{fig:final-model-qp-tip-gradient-focus-spectra}, we
computed the susceptibility at \(\Delta_G=0\) in second-order perturbation theory. Writing $H(\Delta_G)=H_0+\Delta_G V_G+\cdots$, the leading curvature
\begin{equation}\label{eq:susc}
\chi_a^{(2)}
=\sum_{b\ne a}
\frac{|\langle b|V_G|a\rangle|^2}{E_a-E_b},    
\end{equation}
is plotted in Figure~\ref{fig:final-model-qp-gradient-susceptibility}. For each eigenstate $|a\rangle$, we plot $-(\chi_a^{(2)}-\chi_0^{(2)})$, which shows the susceptibility of that level to drop towards the ground state upon increasing $\Delta_G$.
Positive values, shown in red, indicate that the excitation gap is expected to
decrease to quadratic order in \(\Delta_G\); negative values indicate that the
state is pushed up relative to the ground state.  The circled states are the states previously identified to undergo avoided crossing. The susceptibility also reveals some states that are close in energy but do not play an important role in the avoided crossing, although they may be expected to do so based on perturbation theory. For example, at \(N_e=6\), the blue state at \(L_z=5\) -- which is also the one with largest local charge difference compared to the ground state [cf. the bright yellow state in \Cref{fig:qp_lz_spectra}(a)] -- drifts slowly upwards upon turning on $\Delta_G$, in contrast to the circled red state which is the one responsible for the avoided crossing. A similar pattern repeats consistently in other system sizes in Figure~\ref{fig:final-model-qp-gradient-susceptibility}.

Naively, one may expect that the states undergoing avoided crossings and therefore causing the LDOS splitting may be related to different amount of charge trapped near the impurity. For example, in the quasihole case, it was shown that the ground state has three $e/3$ quasiholes piled up around the impurity, while the first excited states have two $e/3$ quasiholes at the impurity and the third $e/3$ futher away~\cite{Wagner26}. These states, with different amounts of trapped charge, naturally have different $L_z$ values and one may expect them to mix once we turn on $\Delta_G$, resulting in LDOS splitting. This intuitive picture turns out to be too simplistic in the FQH case in general. For example, in Figure~\ref{fig:final-model-qp-tip-gradient-focus-spectra} it is evident that the ground state and first excited states are not sharply  distinguished by the amount of trapped charge (although they \emph{do} have different $L_z$ values).

Thus, we conclude that LDOS splitting is indeed related to the breaking of rotation symmetry (via $\Delta_G$ in our model), which causes avoided crossing between states with different $\langle L_z\rangle$ character. The relevant states, however, do not necessarily need to have different amount of charge trapped at the impurity -- due to the large number of possible configurations in the many-body spectrum, the states undergoing avoided level crossing can have similar $\langle Q_A\rangle$ values near the impurity, but differ in charge configurations \emph{farther away} from the impurity.  To demonstrate this more explicitly,  we compute the integrated excess
charge in a spherical cap defined by angle $\theta$ around the impurity:
\begin{equation}\label{eq:Qexc}
Q_{\mathrm{exc}}(\theta)
= \int_0^\theta \mathrm{d} \theta^\prime \int \mathrm{d} \phi^\prime \, \left(\langle \hat \rho(\theta^\prime,\phi^\prime) \rangle - \bar{\rho} \right)\, , \quad   \hat{\rho}(\theta,\phi) \equiv \hat{\Psi}^\dagger(\theta,\phi)\hat \Psi(\theta,\phi)\, ,
\end{equation}
where  $\bar{\rho}$ is the uniform density of the FQH vacuum. 

\begin{figure}[!tb]
\centering
\includegraphics[width=0.31\textwidth]{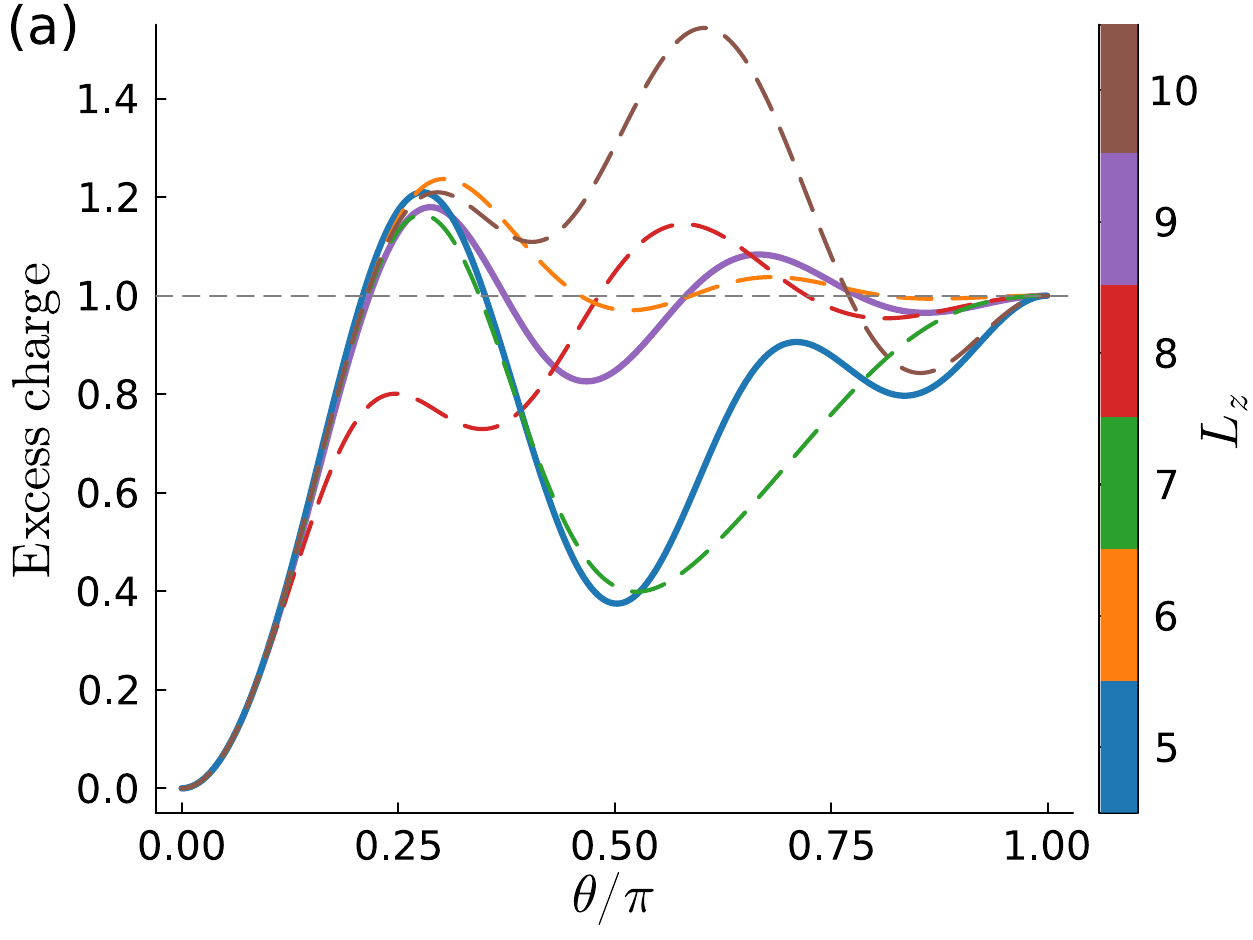} 
\hspace{0.01\textwidth}
\includegraphics[width=0.31\textwidth]{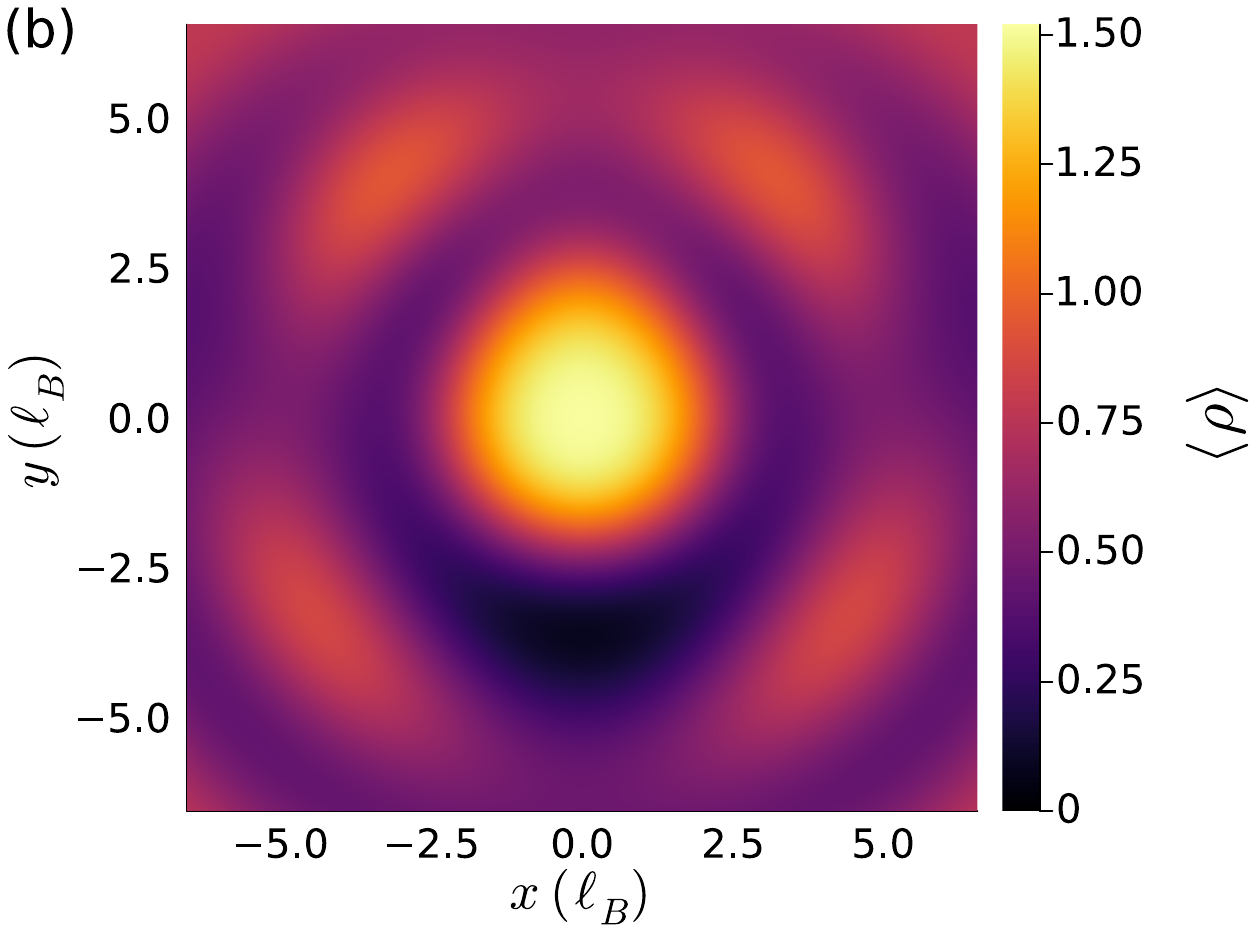}
\hspace{0.01\textwidth}
\includegraphics[width=0.31\textwidth]{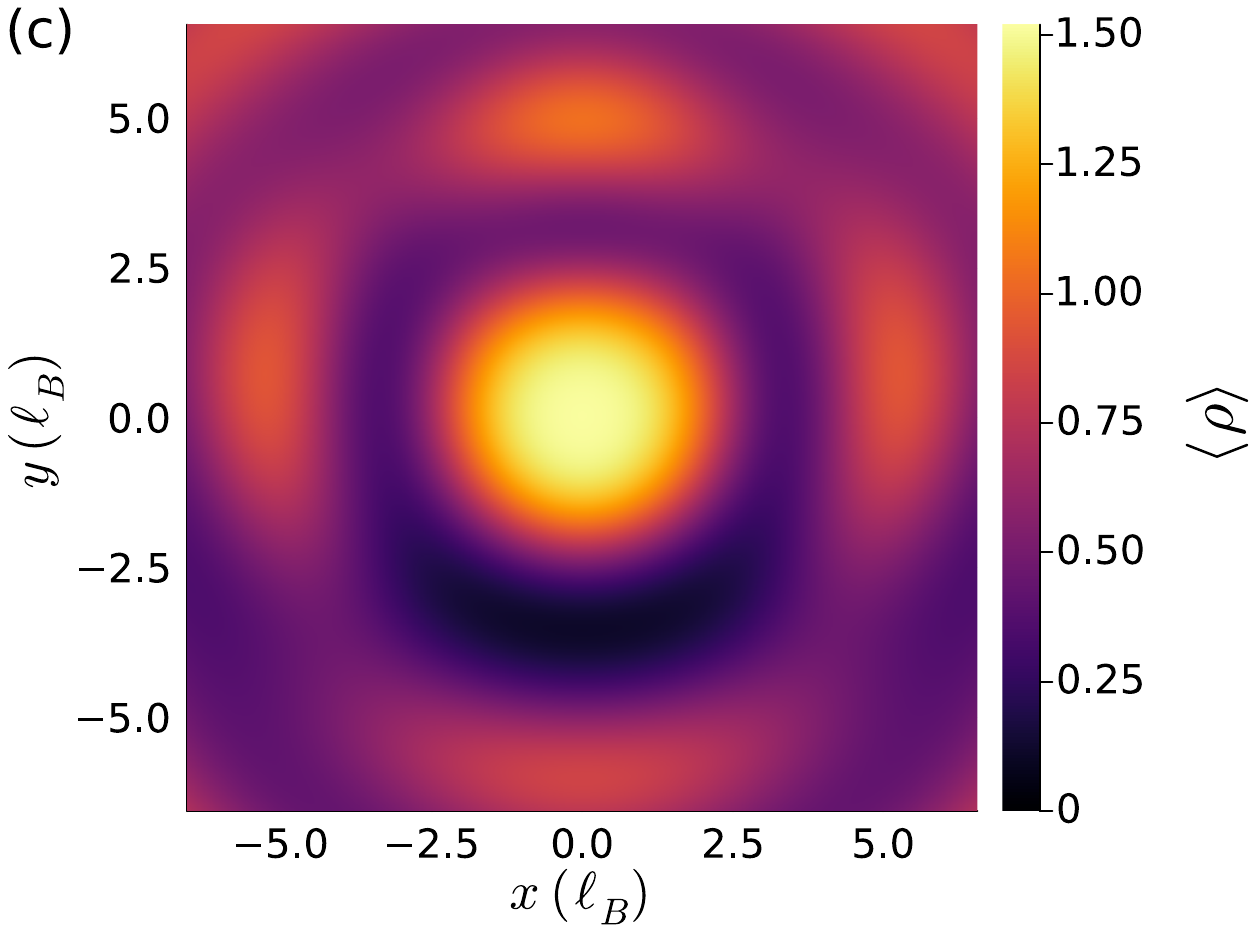}
\caption{
(a): Integrated excess charge, \(Q_{\mathrm{exc}}(\theta)\) in Eq.~(\ref{eq:Qexc}), for the isotropic ground states in different angular momentum sectors. The interaction is tuned to $d_t = 0.2\,\ell_B$ and $\Delta_G=0$, and the system size is $(N_e, 2Q)=(8,18)$. While some of the excited states can have quite different charge distributions compared to the ground state with $L_z=9$, the excited state with $L_z=5$ -- which is the one relevant for the LDOS splitting, as identified in \Cref{fig:final-model-qp-gradient-susceptibility} -- has nearly identical charge profile in the vicinity of the impurity. (b)-(c): The real-space charge distributions of the ground state [panel (b)] and first excited state [panel (c)] in the presence of anisotropic potential gradient $\Delta_G = 0.27$ (value that corresponds to the avoided crossing as indentified in \Cref{fig:delta_g_extrapolation}). While a full electron remains trapped at the impurity even in the presence of anisotropy (making the densities indistinguishable on scales $r\lesssim 2\ell_B$), the surrounding rings further out develop additional structures that reveal a difference between the two states. 
}
\label{fig:excess_plus_density}
\end{figure}

Figure~\ref{fig:excess_plus_density} shows the excess charge for states in different $L_z$ sectors which include the ground state and lowest-lying excited states. Here, we focus on $N_e = 8$ electrons, where the complete LDOS calculation is not accessible in numerics, but the larger system size allows for a better visualization of the charge profile, and directly applies to smaller systems as well. 
From Figure~\ref{fig:final-model-qp-tip-gradient-focus-spectra} it is clear that the relevant states for LDOS splitting are in $L_z=9$ (ground state) and $L_z=5$ sectors; these states, however, have nearly identical excess charge distributions and only differ in charge profiles farther away from the impurity. This is perhaps unsurprising given that the impurity potential is very strong in this case. This implies that several lowest-lying states have the same charge distribution around the impurity, which locally optimizes their energy with respect to the impurity potential, while  further away from the impurity, the states still differ in their charge structure. Furthermore, the fact that the mixing states have $L_z$ values whose difference grows with system size (Figure~\ref{fig:qp_lz_spectra}) suggests that the LDOS splitting cannot be captured in perturbation theory at finite order due to the selection rule $\Delta m =\pm 1$ obeyed by the dipolar impurity potential [see the comment below Eq.~(\ref{eq:gradient})]. For these two relevant states, 
\Cref{fig:excess_plus_density} also shows the charge density at the anisotropic point that corresponds to the avoided crossing. While at the impurity the charge distribution is largely unaffected, different charge distributions emerge in the surrounding charge rings. It would be interesting to find a direct way of relating this property of the ground state (before the tunnelling action of the tip) to the multiple LDOS resonances observed in experiment.

\section{Additional experimental data}
\subsection{Characterization of the device \& IQH and FQH states}

Our experiments are performed on pristine monolayer graphene surfaces that typically show no detectable surface impurities over areas of at least $\SI{200}{\nano\meter}\times\SI{200}{\nano\meter}$, with minimal corrugation below $\SI{100}{\pico\meter}$. Such clean and flat surfaces are essential for isolating the effects of individual charged impurities on anyons. The $dI/dV$ spectra, measured as a function of bias voltage $V_B$ and gate voltage $V_G$, reveal Jain-sequence FQH gaps up to high-denominator fractions, showing high quality of the sample as well as measurement conditions. Within each sector of the fourfold $N=0$ Landau level, the fractional gaps appear in a nearly symmetric manner, for example between conjugate fillings such as $1/3$ and $2/3$, and similar behavior is observed across sectors related by $\nu=0$.

\begin{figure}[htbp]
\centering
\includegraphics[width=0.32\textwidth]{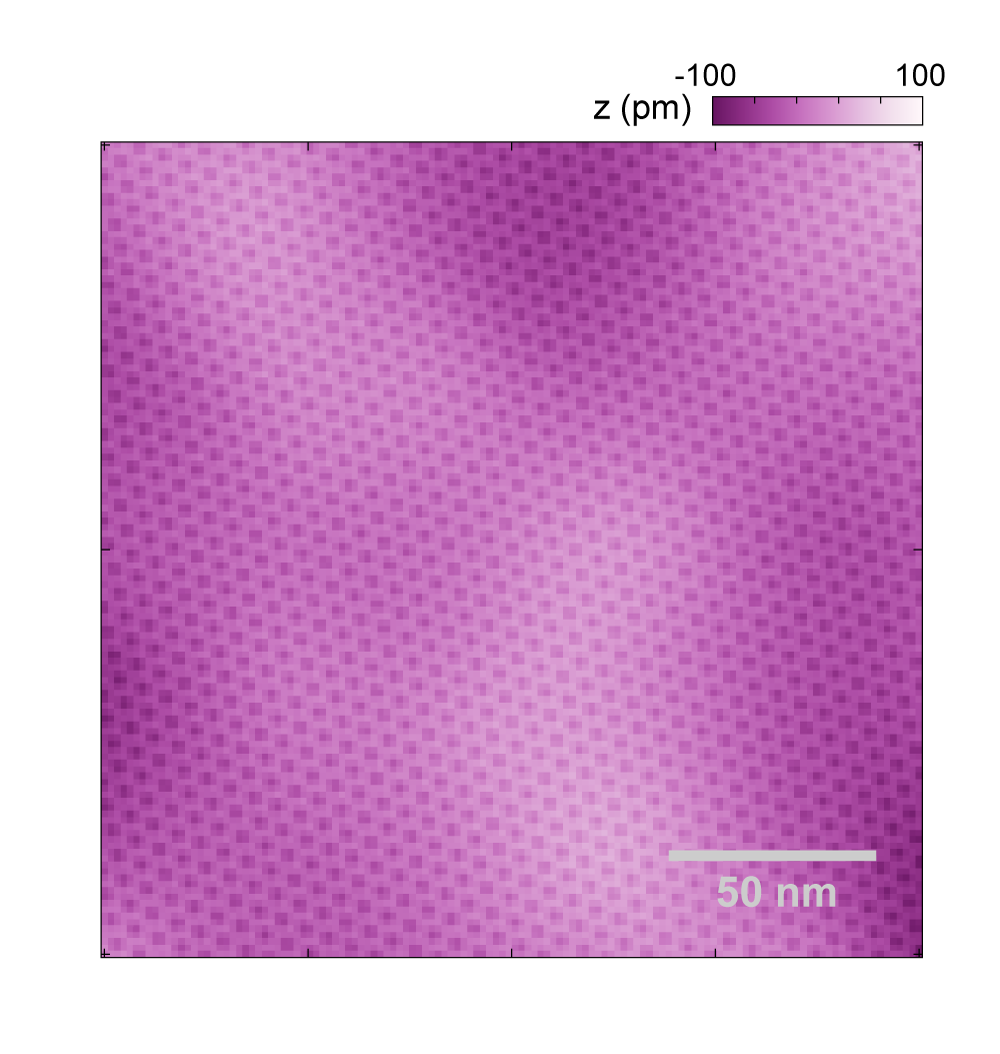}\hfill
\caption{STM topography of a pristine monolayer graphene surface, showing atomically clean regions without defects.}
\label{fig:surface}
\end{figure}

\begin{figure}[htbp]
\centering
\includegraphics[width=\textwidth]{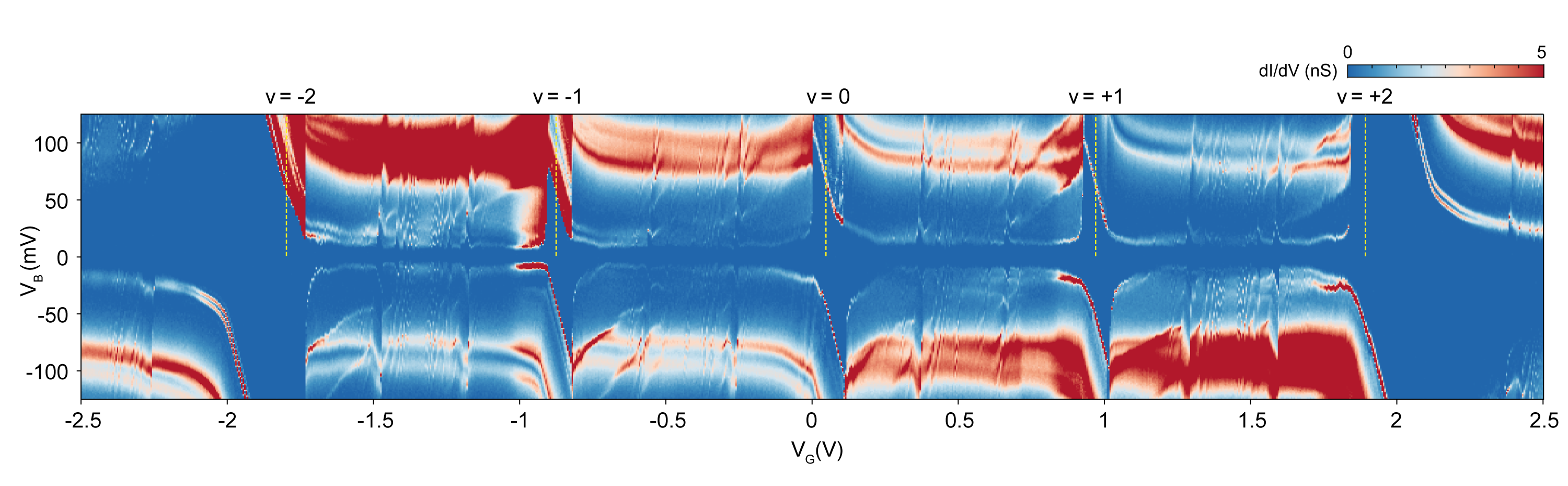}\hfill
\caption{Full gate voltage range spanning the quantum Hall gaps from $\nu=-2$ to $\nu=2$. Data taken at $B=\SI{13.9}{\tesla}$.}
\label{fig:IQH}
\end{figure}

\begin{figure}[htbp]
\centering
\includegraphics[width=0.7\textwidth]{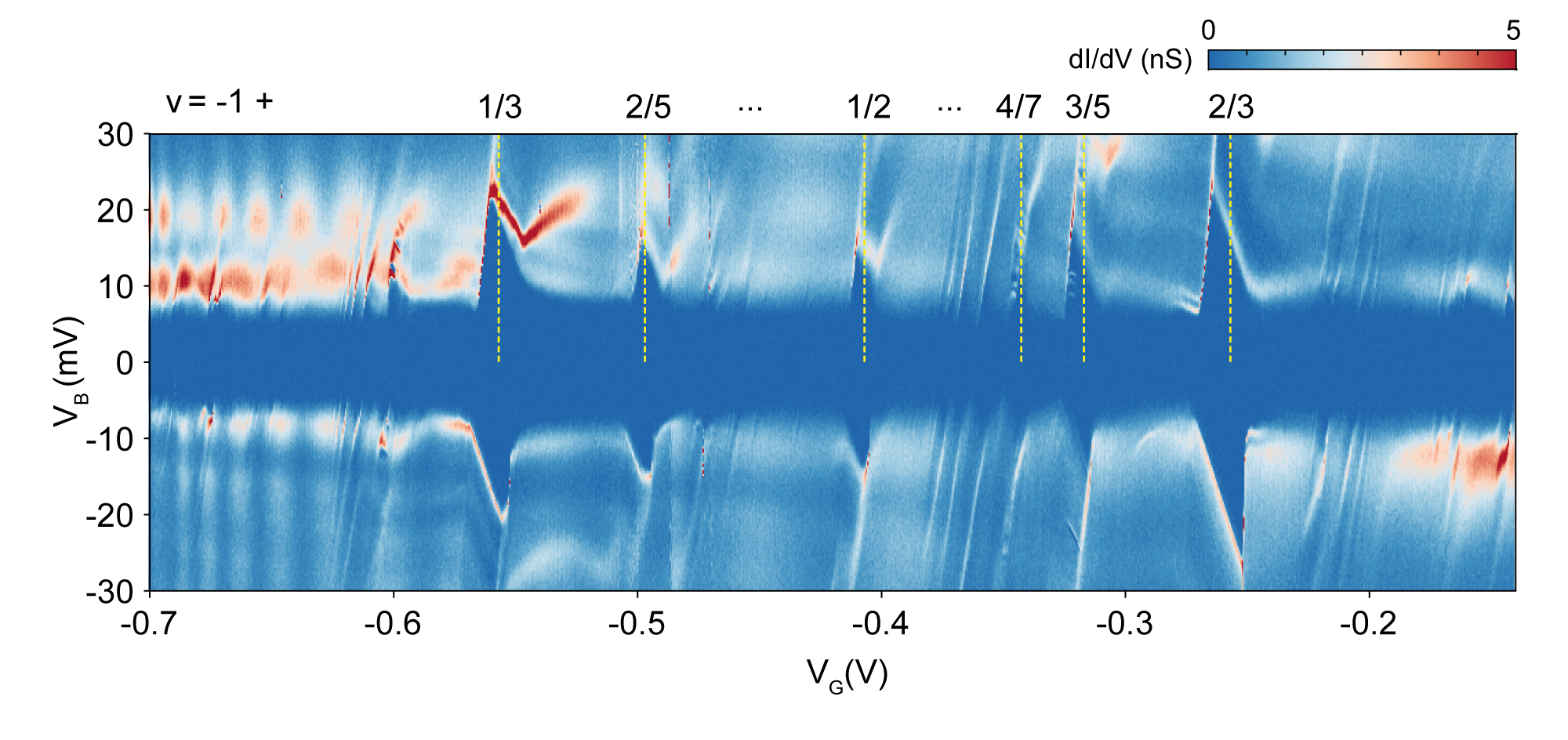}\hfill
\caption{Zoom-in of the LL sector between $\nu=-1$ to $\nu=0$, showing the even-denominator gap at $\nu=-1/2$, together with the standard Jain sequence. The oscillatory feature in the range between $V_G = \SI{-0.7}{\volt}$ to $\SI{-0.6}{\volt}$ is atomic drift. Data taken at $B=\SI{13.9}{\tesla}$.}
\label{fig:FQH1}
\end{figure}

\begin{figure}[htbp]
\centering
\includegraphics[width=0.7\textwidth]{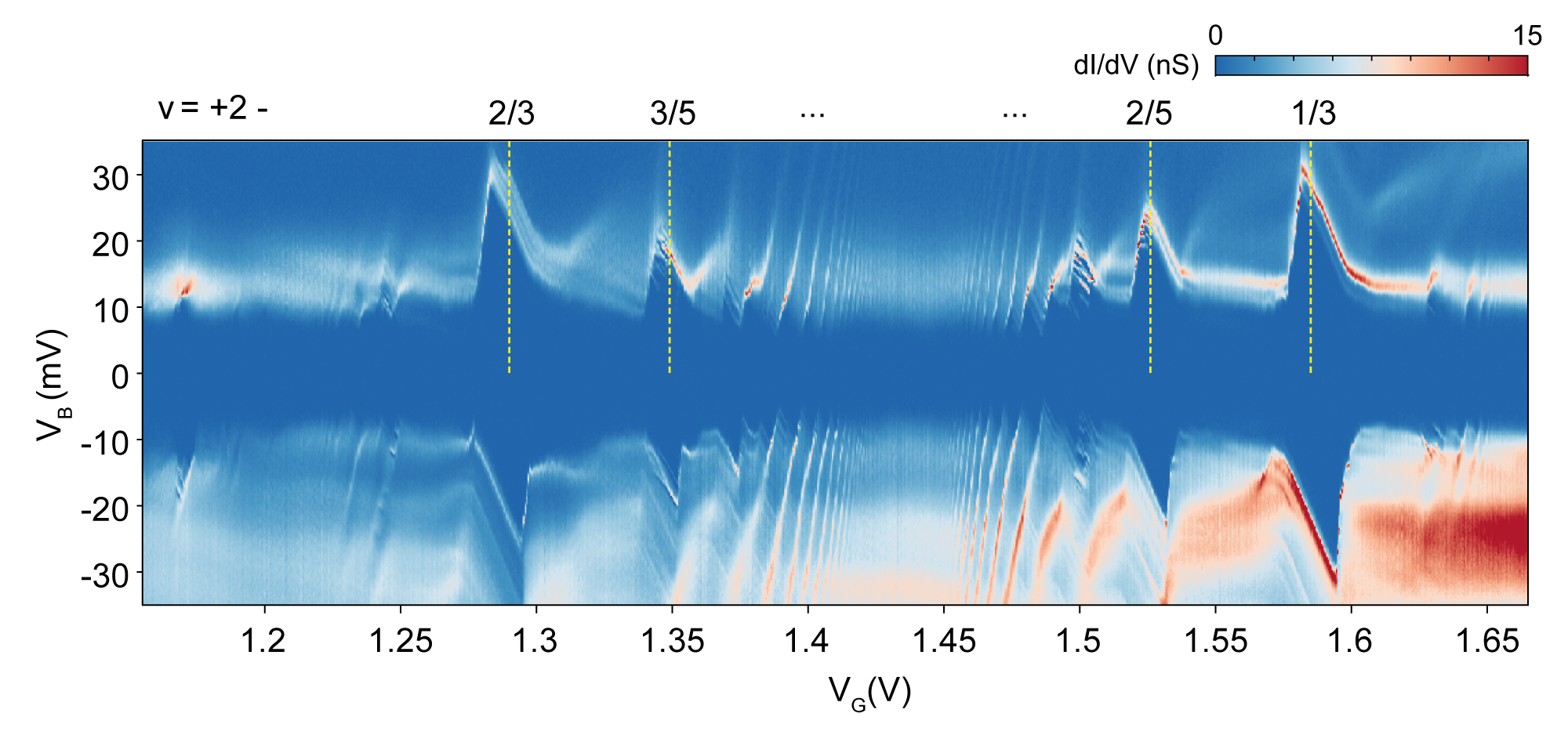}\hfill
\caption{Zoom-in of the LL sector between $\nu=1$ to $\nu=2$, showing symmetric Jain sequence as in Fig. 1a. Data taken at $B=\SI{13.9}{\tesla}$.}
\label{fig:FQH2}
\end{figure}

\FloatBarrier

\subsection{Spectra for different impurities}

We performed spectroscopy near several charged impurities, labeled $A$-$D$, where impurity $A$ corresponds to the data shown in the Main Text. We find that the splitting of the lowest level, observed only inside the FQH gap near the defect potential, is reproducibly seen across different fractional states and for different charged impurities. We also show an example of impurity $E$, which does not appear to be charged and therefore does not trap quasiparticles.

\begin{figure}[htbp]
\centering
\includegraphics[width=0.7\textwidth]{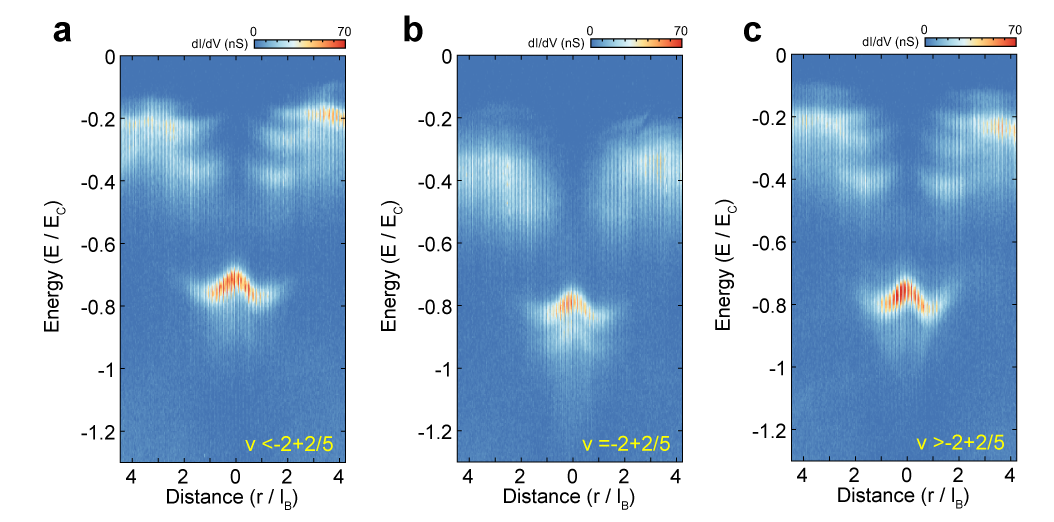}\hfill
\caption{Energy level splitting at the $\nu=-2+2/5$ state. Spatially resolved tunneling spectra showing $dI/dV$ as a function of $V_B$ and distance from the defect at gate voltages corresponding to $nu<-2+2/5$ (a), $\nu=-2+2/5$ (b), and $\nu>-2+2/5$ (c). The splitting is well resolved only inside the fractional quantum Hall gap. Data were taken along a linecut passing $\SI{3}{\nano\meter}$ from the defect center. Impurity C.}
\label{fig:EDF2}
\end{figure}

\begin{figure}[htbp]
\centering
\includegraphics[width=0.45\textwidth]{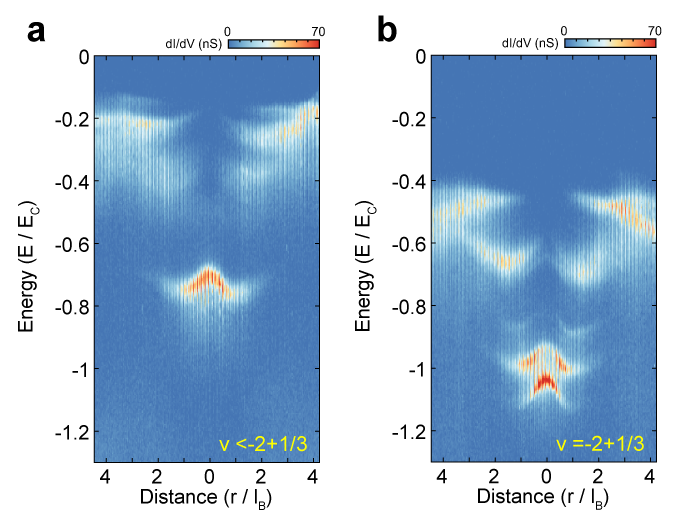}\hfill
\caption{Energy level splitting at the $\nu=-2+1/3$ state for a different tip and impurity potential. Spatially resolved tunneling spectra showing $dI/dV$ as a function of $V_B$ and distance from the defect at gate voltages corresponding to $\nu<-2+1/3$ (a) and $\nu=-2+1/3$ (b). The $\nu>-2+1/3$ regime is shown in \Cref{fig:EDF2}. The splitting is reproducibly well resolved only inside the fractional quantum Hall gap. Data were taken along a linecut passing $\SI{3}{\nano\meter}$ from the defect center. Impurity C.}
\label{fig:EDF3}
\end{figure}

\begin{figure}[htbp]
\centering
\includegraphics[width=0.65\textwidth]{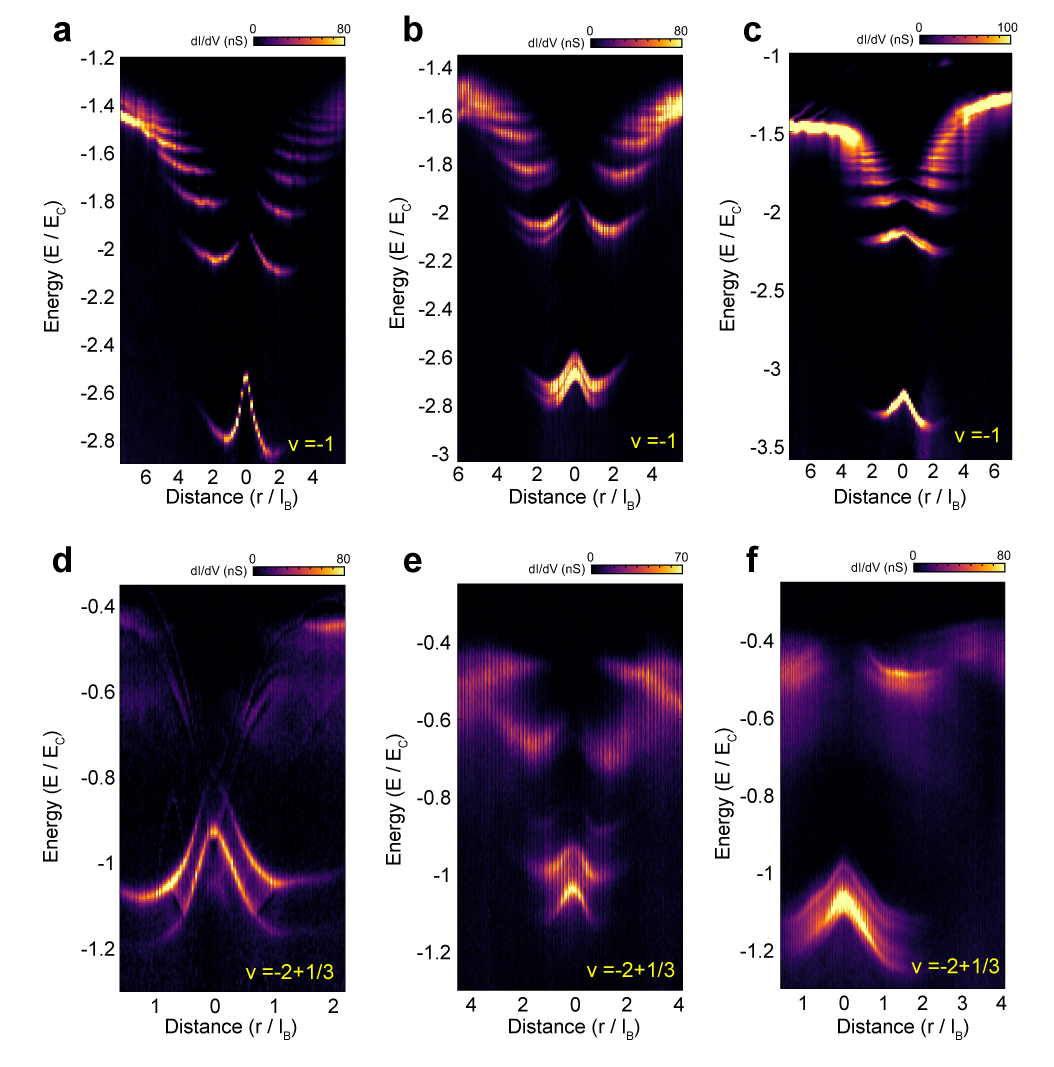}\hfill
\caption{Comparison of different impurity potentials. (a-c) Linecut spectra at $\nu=-1$ for impurity B (subsurface defect; linecut across the center), C (surface defect; linecut $\SI{3}{\nano\meter}$ away from center), and D (surface defect; linecut $\SI{3}{\nano\meter}$ away from center), respectively. (d-f) Linecut spectra at $\nu=-2+1/3$ for impurity B, C, and D, respectively. Impurity B does not exhibit additional splitting for IQH gap at $\nu=-1$. Impurity C shows additional splitting of m = 0 level for the IQH gap at $\nu=-1$, which could be due to spin or valley splitting. Impurity D shows additional splitting for higher $m$ levels at $\nu=-1$. All impurities shown here exhibit splitting at $\nu=-2+1/3$, independent of the defect being surface or subsurface, as well as the presence or absence of additional splitting for IQH gap.}
\label{fig:EDFs}
\end{figure}

\begin{figure}[htbp]
\centering
\includegraphics[width=0.25\textwidth]{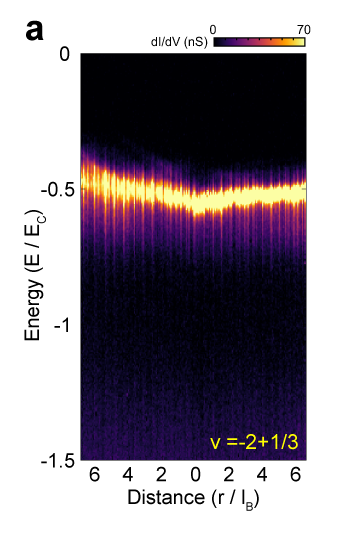}\hfill
\caption{Impurity with shallow/neutral potential. (a) Linecut spectrum at $\nu=-2+1/3$ shows that the impurity does not trap particles. Impurity E (surface defect; linecut $\SI{2}{\nano\meter}$ away from center).}
\label{fig:EDF5}
\end{figure}

\FloatBarrier

\bibliography{biblio_fqhe}